\documentclass[]{nature_mod}
\usepackage{amssymb,amstext,amsmath}
\usepackage{graphicx}
\usepackage{xcolor}
\usepackage{float}
\usepackage{cite}
\usepackage{subcaption}
\usepackage{hyperref}
\usepackage{caption}
\usepackage{dcolumn}
\usepackage{bm} 
\usepackage{multirow}
\usepackage{bigstrut}
\usepackage{makecell,rotating}
\usepackage{mathrsfs}
\usepackage{booktabs}
\usepackage{threeparttable}
\usepackage{chngpage}
\usepackage{float}
\usepackage{color}
\usepackage{amsmath}
\usepackage{caption}
\usepackage{algorithm} 
\usepackage{algpseudocode}   

\usepackage{xr}
\usepackage{color}
\definecolor{darkred}{rgb}{0.75,0,0}
\definecolor{darkgreen}{rgb}{0,0.5,0}
\definecolor{darkblue}{rgb}{0,0,0.75}
\definecolor{darkorange}{rgb}{1,0.9,0.1}
\definecolor{dark}{rgb}{0,0,0}

\begin{document}
	\title{Data-driven functional state estimation of complex networks}

\author{Yuan Zhang\textsuperscript{1\#*}, Ziyuan Luo\textsuperscript{1\#}, Wenxuan Xu\textsuperscript{1\#}, Jiayu Wu\textsuperscript{1}, Wenqi Cao\textsuperscript{2}, Ranbo Cheng\textsuperscript{1},  
Tingting Qin\textsuperscript{2}, 
Yuanqing Xia\textsuperscript{1*}, Mohamed Darouach\textsuperscript{3}, Aming Li\textsuperscript{2,4*}, and Tyrone Fernando\textsuperscript{5*}}  

\date{\today}
\maketitle

\begin{enumerate}
  \item School of Automation, Beijing Institute of Technology, Beijing 100081, China
     \item Center for Systems and Control, Peking University, Beijing 100871, China
           \item  Centre de Recherche en Automatique de Nancy, IUT de Longwy, Universit\'{e} de Lorraine, 54400 Cosnes et Romain, France
  \item Center for Multi-Agent Research, Institute for Artificial Intelligence, Peking University, Beijing 100871, China
  \item Department of Electrical Electronic and Computer Engineering, University of Western Australia 6009, Crawley, Australia
   \item[*]Corresponding authors. Emails: zhangyuan14@bit.edu.cn, xia\_yuanqing@bit.edu.cn, amingli@pku.edu.cn, tyrone.fernando@uwa.edu.au
    \item[\text{\#}]These authors contributed equally.
\end{enumerate}

	\begin{abstract}
    The internal state of a dynamical system—a set of variables that defines its evolving configuration—is often hidden and cannot be fully measured, posing a central challenge for real-time monitoring and control. While observers are designed to estimate these latent states from sensor outputs, their classical designs rely on precise system models, which are often unattainable for complex network systems. Here, we introduce a data-driven framework for estimating a targeted set of state variables, known as functional observers, without identifying the model parameters. We establish a fundamental functional observability criterion based on historical trajectories that guarantees the existence of such observers.  We then develop methods to construct observers using either input-output data or partial state data. These observers match or exceed the performance of model-based counterparts while remaining applicable even to unobservable systems. The framework incorporates noise mitigation and can be easily extended to nonlinear networks via Koopman embeddings. We demonstrate its broad utility through applications including sensor fault detection in water networks, load-frequency control in power grids, and target estimation in nonlinear neuronal systems. Our work provides a practical route for real-time target state inference in complex systems where models are unavailable.
	\end{abstract}
	

The internal state of a dynamical system—a set of variables that fully defines its evolution over time, like metabolite concentrations in a cell\cite{pickard2025dynamic,liu2013observability,maclaren2025observing}—is essential for tasks ranging from feedback stabilization and status monitoring to performance optimization\cite{simon2006optimal,chen2012robust}. However, direct measurement of the full state is often impossible in large-scale networks due to physical or economic constraints\cite{Fixed_Mode,Stefano2012Stability}. Observers, dynamical systems designed to estimate latent states from available output measurements, were introduced to address this gap, with the Luenberger observer establishing the foundational theory for linear systems\cite{luenberger1966observers,luenberger1971introduction}. In many modern applications, from power grids to biological networks, estimating the entire high-dimensional state is neither necessary nor efficient\cite{liu2013observability,montanari2022functional,pickard2025dynamic}. Instead, the focus has shifted to \emph{functional observers}—leaner estimators that reconstruct only a targeted linear function of the state (e.g., critical node subsets or aggregated variables of target nodes), thereby reducing computational burden and system requirements\cite{darouach2000existence,fernando2010functional,maclaren2025observing}, paralleling concepts from target control\cite{gao2014target,klickstein2017energy}.

Conventionally, observer design presupposes complete model knowledge and thus relies on model-based synthesis\cite{luenberger1971introduction,pickard2025dynamic}. Yet, deriving accurate first-principles models for complex, real-world networks is often intractable\cite{yabe2022toward}. Advances in sensing, computation, and storage technologies have enabled the collection of vast, information-rich datasets across diverse domains—from tracking intricate biological processes\cite{turk2013functional,pickard2025dynamic} to managing intelligent infrastructures\cite{ushio2018fluctuating} and modeling physical phenomena\cite{raissi2020hidden}. 
 When first-principle modeling is impractical or costly, this offers a complementary pathway: leveraging the abundance of operational data to understand, predict, and control real-world networked systems\cite{marx2013big,de2019formulas,baggio2021data,yabe2022toward}.  This has spurred the development of data-driven control, which broadly falls into two categories\cite{yeung2019learning,coulson2019data,baggio2021data,yabe2022toward}. The first involves a two-step ``identify-then-design'' process\cite{gevers1993towards,van1995identification}, while the second, more direct approach synthesizes controllers straight from data, proving vital when data are insufficient for unique system identification\cite{coulson2019data,de2019formulas,champion2019data,van2020data}. 
 While data-driven control has matured, with rigorous frameworks now established\cite{de2019formulas,baggio2021data}, particularly in predictive control\cite{coulson2019data,berberich2020data,krishnan2021direct,breschi2023data}, its counterpart—data-driven \emph{estimation}—has received less systematic attention.
Existing works on data-driven prediction\cite{markovsky2008data,wu2024predicting,lusch2018deep,yeung2019learning,chang2019neural} focus mainly on short-horizon forecasting outputs rather than estimating latent states in real time. Recent attempts to design full-state observers from data\cite{Turan2021,wolff2024robust,disaro2024equivalence} still require complete historical state information and produce high-order estimators, limiting their applicability to large-scale or unobservable systems. A general, direct, and scalable framework for data-driven functional observer design remains an open challenge.

 In this paper, we develop an interpretable framework for learning functional state estimators for complex networks directly from data. Specifically, we design observers that estimate linear functions of the state without system identification or explicit model parameters. For linear time-invariant systems, we establish a data-driven functional observability criterion based on the rank of Hankel matrices, which provides necessary and sufficient conditions for the existence of functional observers and for uniquely determining the target state from input-output data. This criterion also enables data-driven functional observability analysis and sensor placement in complex networks. Building on this, we develop systematic procedures to learn functional observers directly from data: one yielding minimum-order observers, and another employing an augmentation strategy for reduced-order observers when the minimum order is unattainable. For the latter, we propose two designs based on data length and availability.
Comprehensive evaluations demonstrate that our data-driven functional observers achieve estimation performance comparable to or better than model-based designs, with reduced order and computational cost. We further analyze the effects of network structure, introduce noise mitigation strategies, and demonstrate a natural extension of the framework to nonlinear systems via Koopman operators. Finally, we validate our approach through three case studies: (i) sensor fault detection in water networks, (ii) observer-based frequency regulation in power grids, and (iii) target-state estimation in nonlinear neural networks using limited sensors.

	
	\section*{Modeling framework}
	
	We consider a network system governed by the following linear time-invariant dynamics
	\begin{equation}\label{eq:system}
	\begin{aligned}
	\mathbf{x}(t+1) &=\mathbf{A}\,\mathbf{x}(t) + \mathbf{B}\,\mathbf{u}(t),\\[1mm]
	\mathbf{y}(t)   &= \mathbf{C}\,\mathbf{x}(t),\\[1mm]
	\mathbf{z}(t)   &= \mathbf{F}\,\mathbf{x}(t),
	\end{aligned}
	\end{equation}
	where the vector $\mathbf{x}(t)=[x_1(t),x_2(t),\cdots,x_n(t)]^\intercal \in \mathbb{R}^{n}$ represents the state of the system at time $t$, with $x_i(t)$ the state of node $i$,  $\mathbf{u}(t) \in \mathbb{R}^{m}$ the input vector, and $\mathbf{y}(t)\in \mathbb{R}^{p}$ the output vector. The matrix $\mathbf{A}\in \mathbb{R}^{n\times n}$ captures the weighted adjacency matrix of the network, $\mathbf{B}\in \mathbb{R}^{n\times m}$ and $\mathbf{C}\in \mathbb{R}^{p\times n}$ are, respectively, the input matrix and output matrix. 
    The matrix $\mathbf{F}\in \mathbb{R}^{r\times n}$, called functional matrix,  defines a linear function of states $\mathbf{z}(t)=[z_1(t),z_2(t),\cdots,z_r(t)]^\intercal\in \mathbb{R}^{r}$ that is of interest 
    (functional or target states), which may correspond to a subset of node states  (namely, target nodes), an aggregate (e.g., average value\cite{maclaren2025observing}) over clusters, or the differences between particular nodes\cite{niazi2020average,trinh2011functional}. 
    The outputs (functional states) are called dedicated if each row of $\mathbf{C}$ ($\mathbf{F}$) contains only one nonzero entry. Similarly, the inputs are dedicated if each column of $\mathbf{B}$ contains only one nonzero entry.  Without losing any generality, assume that both $\mathbf{C}$ and $\mathbf{F}$ have full row rank. 
    

	
\begin{figure}
		\centering
\includegraphics[width=0.9\linewidth]{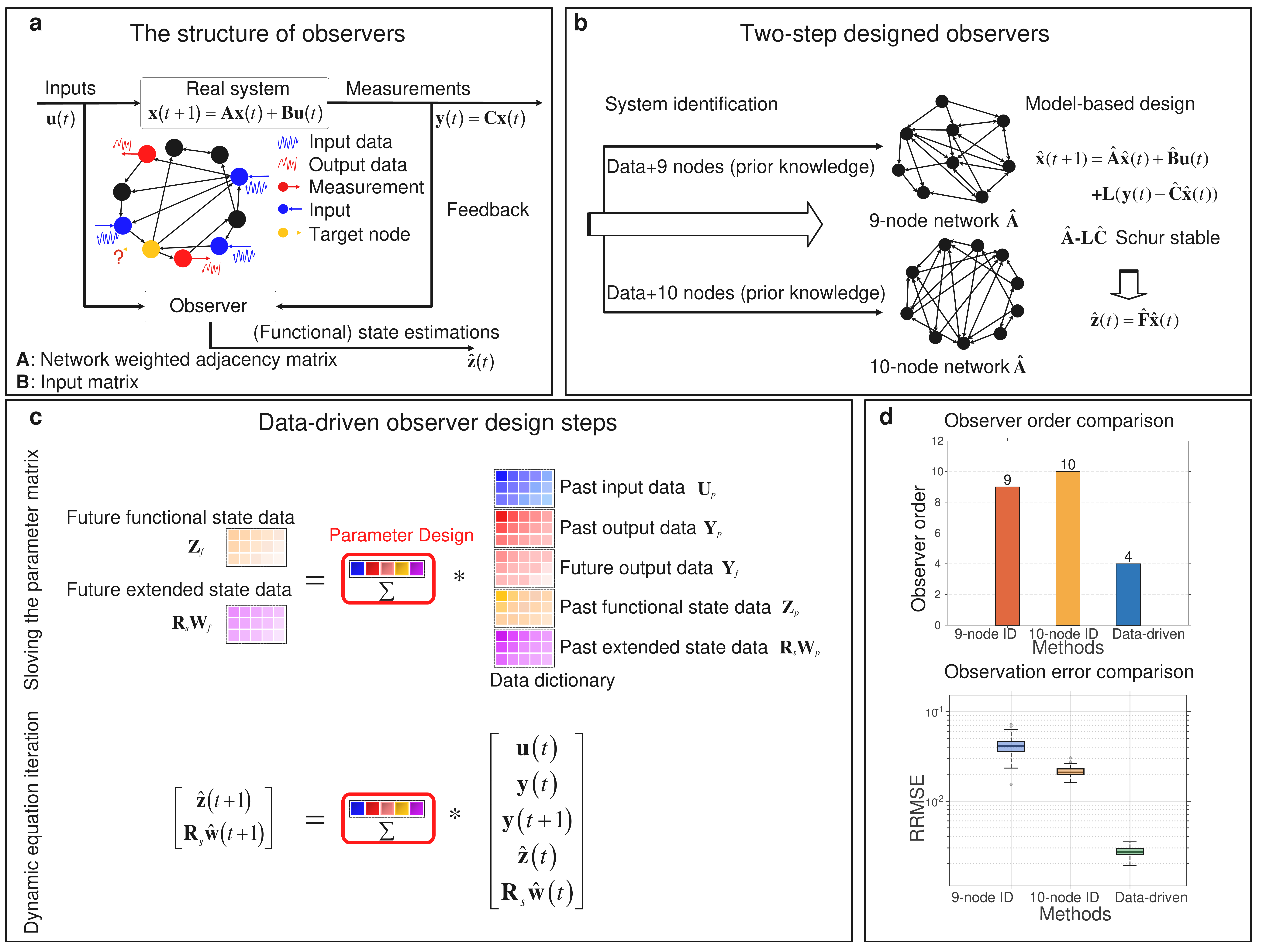}
		\caption{{{\textbf{
     Two-step designed observers versus the proposed direct data-driven functional observer for not fully observable systems.}}     \textbf{a} The general structure of an observer, which uses real-time input and output data to estimate the system's internal state. The illustrative network has \(n=10\) nodes with edge weights randomly drawn from a uniform distribution on \((0,1)\). The system has \(m=3\) inputs (blue nodes), \(p=2\) outputs (red nodes), and a single target functional state \(r=1\) (yellow node). This system is controllable and functionally observable, but not fully state observable (${\rm rank}(\mathbf{O(A,C)})={\rm rank}(\left[\mathbf{O(A,C)^\intercal},\mathbf{O(A,F)^\intercal}\right]^\intercal)= 8 < n = 10$).
\textbf{b} Network topologies identified from a single noisy input-output trajectory (the output is contaminated by independent identically distributed (i.i.d.)  white noise with variance $0.01$) using a subspace-based method (``Methods''), assuming different prior state dimensions (9 and 10), where $({\hat {\mathbf  A}}, \hat {\mathbf B}, \hat {\mathbf C}, \hat {\mathbf F})$ are the identified parameters ($(\hat {\mathbf C}, \hat {\mathbf F})$ is collectively identified by regarding $[\mathbf y^\intercal(t),\mathbf z^\intercal(t)]^\intercal$ as the output). Corresponding Luenberger observers were designed for full-state estimation (see Fig.~S1 for trajectories of the observers).
\textbf{c} The proposed data-driven framework for functional observer design. The observer parameters are learned directly from data and the dynamic equation iteration enables real-time estimation of the target state without intermediate full-state reconstruction.
\textbf{d} Performance comparison. \textit{Top:} Orders of the designed observers. The proposed functional observer is of lower order. \textit{Bottom:} RRMSE of the functional state estimation over 100 independent experiments. Here, the functional observer is constructed from the noisy historical data using IO-data based design detailed subsequently.  The RRMSE was computed over the time period \([50,100]\) to exclude transient effects. For each experiment, the initial state of the original system was randomly initialized, while all observers started from zero.
            }} 
		\label{fig: motivation-framework}
	\end{figure}

	Our objective is to design an observer that produces an estimate $\mathbf{\mathbf{\hat{z}}}(t)$ from the measured input $\mathbf{u}(t)$ and output $\mathbf{y}(t)$ such that the estimation error vanishes $
	\lim_{t\to\infty}\|\mathbf{z}(t)-\mathbf{\mathbf{\hat{z}}}(t)\|=0
	$ asymptotically (see Fig.~\ref{fig: motivation-framework}a for the structure of a general observer).
	Such an estimator is called a functional observer\cite{darouach2000existence,fernando2010functional}. When $\mathbf{F}=\mathbf{I}_n$, this observer reduces to the classical Luenberger observer\cite{luenberger1966observers}.  It is known that when the observability matrix
	\begin{equation}
	\mathbf{O(A,C)}=[\mathbf{C}^\intercal, (\mathbf{C}\mathbf{{ A}})^\intercal, (\mathbf{C}{ \mathbf{A}}^2)^\intercal,\cdots, (\mathbf{C}{\mathbf{A}}^{n-1})^\intercal]^\intercal, 
	\end{equation}
	has full column rank, equivalently, system (\ref{eq:system}) is observable, a Luenberger observer exists\cite{luenberger1971introduction}. However, in many applications, as mentioned previously, it is sufficient and computationally advantageous to estimate only a specific function of the state\cite{montanari2022functional}.
	
	
	A functional observer may exist even if the system~(\ref{eq:system}) is not observable. Indeed, if system (\ref{eq:system}) is functionally observable (see Supplementary Note 1 for the definition)\cite{fernando2010functional,fernando2010functional2}, mathematically, the following condition holds,
	\begin{equation} \label{fun-criterion}
	{\rm rank}\left[\begin{array}{c}\mathbf{O(A,C)}\\
	\mathbf{F}
	\end{array}\right]={\rm rank}(\mathbf{O(A,C)}),
	\end{equation}
	which is also equivalent to ${\rm rank}(\mathbf{O(A,C)})={\rm rank}(\left[\mathbf{O(A,C)^\intercal},\mathbf{O(A,F)^\intercal}\right]^\intercal)$, a functional observer exists so that the estimator error $\mathbf{z}(t)-\mathbf{\hat z}(t)$ can converge asymptotically to zero with an arbitrary rate. While a reduced-order Luenberger observer has an order of $n-p$, the order of a functional observer can be further reduced to some number between $r$ and $n-p$\cite{fernando2010functional,trinh2011functional}. Detailed theoretical conditions for the existence of functional observers are provided in ``Methods'' and Supplementary Note~1.
	Existing approaches for functional observer design rely on precise knowledge of system parameters $(\mathbf{A},\mathbf{B},\mathbf{C},\mathbf{F})$. In practice, implementing these methods requires first-principles modeling or system identification (ID), thus forming a two-step design method (namely, ID followed by model-based design), which can be computationally inefficient and prone to numerical instabilities\cite{markovsky2023data}.  {To see this, we provide an illustrative example in Fig.~\ref{fig: motivation-framework}. In this example, the input-output data of an unobservable network system leads to the identification of two different network topologies given different prior knowledge of the system state dimension (Fig.~\ref{fig: motivation-framework}b). Then, two Luenberger observers are constructed based on the identified models, which have larger orders than the proposed data-driven functional observer (Fig.~\ref{fig: motivation-framework}c,d; detailed in Section ``Data-driven reduced-order functional observer design''). Fig.~\ref{fig: motivation-framework}d indicates that the two-step design approach tends to exhibit larger estimation errors and higher variances than the direct data-driven method. This is perhaps because numerical errors induced in each step may accumulate, leading to unreliable estimation performance. Throughout this paper, to evaluate the performance of an observer during a given period $[T_1,T_2]$, we define the relative root mean square error (RRMSE) as
		\begin{equation} \label{RRMSE}
		{\rm RRMSE}= \frac{\sqrt{\sum_{k=T_1}^{T_2} \| \mathbf{z}_{\rm esti}(k) - \mathbf{z}_{\rm true}(k)\|_2^2}}{\sqrt{\sum_{k=T_1}^{T_2} \|\mathbf{z}_{\rm true}(k)\|_2^2}},
		\end{equation}where $\mathbf{z}_{\rm true}(k)$ is the ground truth and $\mathbf{z}_{\rm esti}(k)$ is the estimated value. The period $[T_1,T_2]$ will be omitted when it is clear from the context.}
	
To address the aforementioned limitations, here we propose a data-driven approach for functional observer design that eliminates dependence on exact model parameters while maintaining performance comparable to or exceeding model-based approaches, and applies naturally to nonlinear systems with the aid of Koopman operators. 
	
	
    \section*{Results}
  {\bf{Data-driven criterion for functional observability.}}  In practice, the system matrices $(\mathbf{A},\mathbf{B},\mathbf{C},\mathbf{F})$ are rarely known exactly. Instead, one typically has access to historical trajectories of system~\eqref{eq:system}, $\{\mathbf{u}_d(t), \, \mathbf{y}_d(t), \, \mathbf{z}_d(t), \, \mathbf{w}_d(t)\}_{t=0}^{T-1}$,
where $\mathbf{U}=\{\mathbf{u}_d(t)\}$ and $\mathbf{Y}=\{\mathbf{y}_d(t)\}$ denote the input and output sequences, $\mathbf{Z}=\{\mathbf{z}_d(t)\}$ the functional state trajectory, and $\mathbf{W}=\{\mathbf{w}_d(t)\}$ an additional (extended) set of linear functions of the state. The horizon length is $T$. The historical sequences $\mathbf{Z}$ and $\mathbf{W}$ may be obtained from costly sensors not available in routine operation, or inferred offline through computationally demanding procedures, making them impractical for online use\cite{wolff2024robust}. Crucially, without historical functional state data $\mathbf{Z}$, asymptotic estimation of $z(t)$ becomes impossible since any invertible coordinate transformation of internal state variables would leave input-output data unchanged\cite{trinh2011functional}. We denote by $\mathbf{X}=\{\mathbf{x}_d(t)\}$ the corresponding latent state trajectory, which is often inaccessible.  Because system~\eqref{eq:system} may not be observable, the available data $\{\mathbf{U}, \mathbf{Y}, \mathbf{Z}, \mathbf{W}\}_{t=0}^{T-1}$ could not suffice to identify the system uniquely, even up to invertible transformations\cite{verhaegen2007filtering} (see ``Methods'').

	Our approach exploits this data by constructing Hankel matrices to capture the system dynamics over finite time windows. For a signal sequence $\mathbf{V}=\{\mathbf{v}_d(t)\}_{t=0}^{T-1}$, with $\mathbf{V}=\mathbf{U},\mathbf{Y},\mathbf{Z}$, and $\mathbf{W}$, the Hankel matrix of order $k$ is defined as
	\begin{equation}\label{eq:hankel}
	\mathbf{V}_{i,k,N} = \begin{bmatrix}
	\mathbf{v}_d(i) & \mathbf{v}_d(i+1) & \cdots & \mathbf{v}_d(i+N-1) \\
	\mathbf{v}_d(i+1) & \mathbf{v}_d(i+2) & \cdots & \mathbf{v}_d(i+N) \\
	\vdots & \vdots & \ddots & \vdots \\
	\mathbf{v}_d(i+k-1) & \mathbf{v}_d(i+k) & \cdots & \mathbf{v}_d(i+k+N-2)
	\end{bmatrix},
	\end{equation}where $k$ indicates the number of block rows, $N$ is the number of columns, and $i$ specifies the initial time index of the matrix. We also write $\mathbf{V}_{i,N}=[\mathbf{v}_d(i),\mathbf{v}_d(i+1),...,\mathbf{v}_d(i+N-1)]$.
	We partition the data into ``past'' (denoted by $\mathbf{V}_p$) and ``future'' ($\mathbf{V}_f$) segments, defined to be $\mathbf{V}_p=\mathbf{V}_{0,T-1}$ and $\mathbf{V}_f=\mathbf{V}_{1,T-1}$,  respectively. Under a standard persistency of excitation (PE) condition 
    (Supplementary Note~2), it can be shown that the system is functionally observable if and only if the following rank condition holds:
	\begin{equation}\label{eq:rank_condition}
	\mathrm{rank}\begin{bmatrix}
	\mathbf{U}_{0,k,T-k+1}\\[1mm]
	\mathbf{Y}_{0,k,T-k+1}\\[1mm]
	\mathbf{Z}_{0,T-k+1}
	\end{bmatrix} = \mathrm{rank}\begin{bmatrix}
	\mathbf{U}_{0,k,T-k+1}\\[1mm]
	\mathbf{Y}_{0,k,T-k+1}
	\end{bmatrix}, \quad k \ge l(\mathbf{A},\mathbf{C}),
	\end{equation}
	where $l(\mathbf{A},\mathbf{C})$ denotes the observability index of the pair $(\mathbf{A},\mathbf{C})$, namely, the smallest $k\in {\mathbb{N}}$ such that ${\rm rank}(\mathbf{O}_k(\mathbf{A},\mathbf{C}))$ reaches ${\rm rank}(\mathbf{O}(\mathbf{A},\mathbf{C}))$, with $\mathbf{O}_k(\mathbf{A},\mathbf{C}) = [\mathbf{C}^{\intercal}, (\mathbf{C}\mathbf{A})^{\intercal}, \dots,(\mathbf{\mathbf{C}}\mathbf{A}^{k-1})^{\intercal}]^{\intercal}$ (if $l(\mathbf{A},\mathbf{C})$ is not known a priori, we can take $k$ to be its upper bound, $n$).  This rank condition ensures that, for a sufficiently long input-output trajectory $\{\mathbf{u}(t),\mathbf{y}(t)\}_{t=0}^{k-1}$ (with trajectory length $k$ satisfying $k\ge l(\mathbf{A},\mathbf{C})$), there is a unique initial functional state $\mathbf{z}(0)$  corresponding to the trajectory, even when the system is not observable (see Supplementary Note 3). In the special case where $\mathbf{F}=\mathbf{I}_n$, Eq. \eqref{eq:rank_condition} provides a data-based necessary and sufficient condition for full state observability. Note that each row of $\mathbf{y}_d(t)$ corresponds to a sensor. Hence, condition \eqref{eq:rank_condition} can also be leveraged to develop data-driven algorithms for sensor placement, enabling (functional) observability in complex networks—an extensively studied problem that traditionally relies on model-based methods\cite{Y.Y.2011Controllability,Yuan2013Exact,liu2013observability,montanari2022functional,zhang2023functional}.

	

	\noindent \textbf{Learning a minimum-order functional observer from data.}
	We first present the construction of a minimum-order functional observer, also known as a Darouach observer\cite{darouach2000existence}, whose order is equal to the dimension $r$ of the functional state, directly from data. Under the PE condition, it can be established (see Supplementary Note~4) that a Darouach observer exists if and only if there exists a matrix 
	$
	\mathbf{\Sigma} = \big[\mathbf{\Sigma}_{\mathbf{U}_p}\ \mathbf{\Sigma}_{\mathbf{Y}_p}\ \mathbf{\Sigma}_{\mathbf{Y}_f}\ \mathbf{\Sigma}_{\mathbf{Z}_p}\big],
	$
	in which $\mathbf{\Sigma}_{\mathbf{Z}_p}$ is Schur stable (i.e., with all eigenvalues within the unit circle), such that the future block of the functional state can be written as
	\begin{equation}\label{eq:darouach}
	{\mathbf{Z}_f} = \underbrace{\begin{bmatrix}\mathbf{\Sigma}_{\mathbf{U}_p} & \mathbf{\Sigma}_{\mathbf{Y}_p} & \mathbf{\Sigma}_{\mathbf{Y}_f} & \mathbf{\Sigma}_{\mathbf{Z}_p}\end{bmatrix}}_{\doteq \mathbf{\Sigma}}
	\begin{bmatrix}{\mathbf{U}_p} \\[1mm] {\mathbf{Y}_p} \\[1mm] {\mathbf{Y}_f} \\[1mm] {\mathbf{Z}_p} \end{bmatrix}.
	\end{equation}
	Here, $\mathbf{\Sigma}_{\mathbf{U}_p}$, $\mathbf{\Sigma}_{\mathbf{Y}_p}$, and $\mathbf{\Sigma}_{\mathbf{Y}_f}$ encode the contributions of the past input, past output, and future output data, respectively, while $\mathbf{\Sigma}_{\mathbf{Z}_p}$ governs the recursive estimation dynamics. With this representation, the online update rule for the functional observer is given by
\begin{equation}\label{eq:observer_update_main}
	\mathbf{\mathbf{\hat{z}}}(t+1) = \mathbf{\Sigma}_{\mathbf{U}_p}\,\mathbf{u}(t) + \mathbf{\Sigma}_{\mathbf{Y}_p}\,\mathbf{y}(t) + \mathbf{\Sigma}_{\mathbf{Y}_f}\,\mathbf{y}(t+1) + \mathbf{\Sigma}_{\mathbf{Z}_p}\,\mathbf{\mathbf{\hat{z}}}(t).
	\end{equation}
	This recursion ensures that, starting from any initial condition, the estimation error $\|\mathbf{z}(t)-\mathbf{\mathbf{\hat{z}}}(t)\|_2$ converges to zero asymptotically, where the convergence rate is determined by eigenvalues of $\mathbf{\Sigma}_{\mathbf{Z}_p}$. Note that a general solution for $\mathbf{\Sigma}$ satisfying Eq. \eqref{eq:darouach} is \begin{equation}\label{general-solution}\mathbf{\Sigma}=\mathbf{Z}_f\mathbf{D}^{\dagger}+\mathbf{K}(\mathbf{I}-\mathbf{D}\mathbf{D}^\dagger),\end{equation}where $\mathbf{D}\doteq \left[{\mathbf{U}_p}^{\intercal}, {\mathbf{Y}_p}^{\intercal}, {\mathbf{Y}_f}^{\intercal}, {\mathbf{Z}_p}^{\intercal}\right]^{\intercal}$, $\mathbf{D}^{\dagger}$ takes the Moore-Penrose pseudoinverse of $\mathbf{D}$, and $\mathbf{K}$ is an arbitrary matrix (here, $\mathbf{I}-\mathbf{D}\mathbf{D}^\dagger$ can be replaced by any basis matrix of the left null space of $\mathbf{D}$). Partition  ${\mathbf{D}}^{\dagger}=[\mathbf{H}_{\mathbf{U}_p},\mathbf{H}_{\mathbf{Y}_p} , \mathbf{H}_{\mathbf{Y}_f} , \mathbf{H}_{\mathbf{Z}_p}]$ and  $\mathbf{I}-\mathbf{D}\mathbf{D}^\dagger=[\mathbf{\Delta}_{\mathbf{U}_p},\mathbf{\Delta}_{\mathbf{Y}_p}, \mathbf{\Delta}_{\mathbf{Y}_f},\mathbf{\Delta}_{\mathbf{Z}_p}]$ in accordance with $\mathbf{\Sigma} = \big[\mathbf{\Sigma}_{\mathbf{U}_p}\ \mathbf{\Sigma}_{\mathbf{Y}_p}\ \mathbf{\Sigma}_{\mathbf{Y}_f}\ \mathbf{\Sigma}_{\mathbf{Z}_p}\big]$. Then, the existence of a Schur stable $\mathbf{\Sigma}_{\mathbf{Z}_p}$ satisfying Eq. \eqref{eq:darouach} can be checked by verifying the detectability of $(\mathbf{Z}_f \mathbf{H}_{\mathbf{Z}_p}, \mathbf{\Delta}_{\mathbf{Z}_p})$. Moreover, if $(\mathbf{Z}_f \mathbf{H}_{\mathbf{Z}_p}, \mathbf{\Delta}_{\mathbf{Z}_p})$ is observable, the estimation error can converge to zero at an arbitrary rate by choosing feasible $\mathbf{K}$.  The derivation and stability analysis of Eq.~\eqref{eq:observer_update_main} are provided in Supplementary Note~4.

 	\noindent \textbf{Data-driven reduced-order functional observer design.}
	In some scenarios, a Darouach observer may not exist because the functional state $\mathbf{z}(t)$ is embedded in a higher-order dynamic system and it is impossible to decouple $\mathbf{z}(t)$ from this system with the knowledge of $\mathbf{y}(t)$. To address this, inspired by ref.\cite{fernando2010functional}, we propose an augmentation strategy by defining an augmented functional state:
	\[
	\mathbf{z}'(t) = \begin{bmatrix} \mathbf{z}(t) \\ \mathbf{R}_s\,\mathbf{w}(t) \end{bmatrix},
	\]
	where $\mathbf{R}_s$ is a matrix chosen to extract additional state information from the extended linear function of state $\mathbf{w}(t)$. Note that $\mathbf{w}(t)$ corresponds to the extended functional state data $\mathbf{w}_d(t)$ and the corresponding available data matrix $\mathbf{W}=\{\mathbf{w}_d(t)\}_{t=0}^{T-1}$. The design of $\mathbf{R}_s$ is based on ensuring that the following rank conditions are met:
\begin{align}
	\operatorname{rank}\begin{bmatrix}
	{\mathbf{U}_p} \\[1mm]
	{\mathbf{Y}_p} \\[1mm]
	{\mathbf{Y}_f} \\[1mm]
	{\mathbf{Z}_p} \\[1mm]
	{\mathbf{R}_s}\,{\mathbf{W}_p} \\[1mm]
    {\mathbf{Z}_f} \\[1mm]
	{\mathbf{R}_s}\,{\mathbf{W}_f}
	\end{bmatrix} &= \operatorname{rank}\begin{bmatrix}
	{\mathbf{U}_p} \\[1mm]
	{\mathbf{Y}_p} \\[1mm]
	{\mathbf{Y}_f} \\[1mm]
	{\mathbf{Z}_p} \\[1mm]
	{\mathbf{R}_s}\,{\mathbf{W}_p}
	\end{bmatrix}, \label{condnew1}\\[1mm]
	\operatorname{rank}\begin{bmatrix}
	{\mathbf{U}_p} \\[1mm]
	{\mathbf{Y}_p} \\[1mm]
	{\mathbf{Y}_f} \\[1mm]
	{\mathbf{Z}_p} \\[1mm]
	{\mathbf{R}_s}\,{\mathbf{W}_p}
	\end{bmatrix} &= \operatorname{rank}\begin{bmatrix}
	{\mathbf{U}_p} \\[1mm]
	{\mathbf{Y}_p} \\[1mm]
	{\mathbf{Y}_f} \\[1mm]
	\lambda\begin{bmatrix} {\mathbf{Z}_p} \\[1mm] {\mathbf{R}_s}\,{\mathbf{W}_p} \end{bmatrix} - \begin{bmatrix} {\mathbf{Z}_f} \\[1mm] {\mathbf{R}_s}\,{\mathbf{W}_f} \end{bmatrix}
	\end{bmatrix},\quad \forall\, |\lambda|\ge 1. \label{condnew1-add}
	\end{align} 
 These conditions are equivalent to satisfying condition \eqref{eq:darouach} with ${\mathbf{Z}}_p$ replaced by $[{\mathbf{Z}}_p^\intercal, ({\mathbf{R}_s}{\mathbf{W}}_p)^\intercal]^\intercal $ and ${\mathbf{Z}}_f$ replaced by $[{\mathbf{Z}}_f^\intercal, ({\mathbf{R}_s}{\mathbf{W}}_f)^\intercal]^\intercal$ (see Supplementary Note 4).   
In the case $\mathbf{W}=\mathbf{X}$, i.e., the full state data $\mathbf{X}$ is available, a model-based functional observer of order $n_d$ exists that can track $\mathbf{z}(t)$ asymptotically if and only if there exists an ${\mathbf{R}_s} \in \mathbb{R}^{(n_d-r) \times n}$ such that~\eqref{condnew1}-\eqref{condnew1-add} hold. This establishes the equivalence of the model-based functional observer design and the data-driven one with the full state data. In case no extended functional state data, except for $\mathbf{Z}$, is available,  a subspace-identification inspired method could be used to recover the projection $\mathbf{X}^{\rm proj}$ of $\mathbf{X}$ onto the row space of $\mathbf{O(A,C)}$, provided that the length of input-output (IO) data $T$ is sufficiently large (see ``Methods'' and Proposition 2 in Supplementary Note 4). By replacing $\mathbf{W}$ with $\mathbf{X}^{\rm proj}$, the design for a reduced-order functional observer is still valid. Hence, two approaches can be selected for designing a reduced-order functional observer, depending on the available data and its length. The first approach requires extended functional state data $\mathbf{W}$ beyond $\mathbf{Z}$ (called extended-state based design), but operates with a relatively short data length $T_{\rm extended}$,  with $T_{\rm extended}\ge n+m+1$. In contrast, the second approach does not require extended functional state data (called IO-data based design) but demands a longer data length $T_{\rm IO}$, which should satisfy $T_{\rm IO}\ge T^*_{\rm IO}\doteq n+2(m+1)l(\mathbf{A},\mathbf{C})$ (see ``Data length comparison" in Supplementary Note 4).

	Finding a matrix with the smallest number of rows so that \eqref{condnew1}-\eqref{condnew1-add} hold is challenging. We propose an iterative subspace-intersection approach to find a feasible ${\mathbf{R}_s}$ with rows as small as possible (but without optimality guarantee) so that \eqref{condnew1} and \eqref{condnew1-add} hold (see ``Methods'').  At each step, it derives analytical expressions for ${\mathbf{R}_s} {\mathbf{W}_f}$ to minimize the rank difference between the two sides of \eqref{condnew1}.    Once \eqref{condnew1} holds, \eqref{condnew1-add} reduces to a detectability condition, which is prone to hold in practical cases since
	controllability (a sufficient condition for detectability) is a generic property (see Supplementary Note 4). Once ${\mathbf{R}_s}$ is determined, a reduced-order observer is designed analogously to Eq.~\eqref{eq:observer_update_main}, with the estimate $\mathbf{\hat{z}}(t)$ extracted from the augmented state $\mathbf{\hat{z}}'(t)$.
	When $\mathbf{W}=
    \mathbf{X}^{\rm proj}$ and the system is functionally observable, the worst-case observer order determined by this method is ${\rm rank}({\mathbf{O}}(\mathbf A,\mathbf C))-p$ (Theorem 3 in Supplementary Note 4). This order does not exceed that of a reduced-order Luenberger observer, which, it should be noted, requires full observability of the system—a condition stronger than functional observability.

Fig.~\ref{fig:fig2} compares the performance of our proposed data-driven functional observer against the conventional two-step approach of system identification followed by model-based design (denoted ID+Model-based). Both methods utilize only historical input ($\mathbf{U}$), output ($\mathbf{Y}$), and functional state ($\mathbf{Z}$) data. We employ model-based functional observers as the benchmark rather than Luenberger observers, as the generated systems may not be fully observable. As shown in Figs.~\ref{fig:fig2}a,b, the proposed data-driven design maintains a consistently low RRMSE of approximately $10^{-4}$ across both Erd\H{o}s--R\'enyi (ER) and Barab\'asi--Albert (BA) networks. In contrast, a significant proportion of observers designed via the two-step approach exhibit RRMSE values exceeding $10^{-2}$. This result demonstrates that the data-driven approach achieves superior stability with significantly lower variance in RRMSE performance compared to the traditional method. Notably, this advantage holds consistently for both observable and unobservable systems (see Fig. S2). 
  Fig.~\ref{fig:fig2}c compares the computational time of the two approaches. The two-step approach requires constructing a large-dimensional matrix via the Kronecker product and subsequently computing its inverse (see ``Methods''). In contrast, the data-driven approach directly stacks the input-output-functional state data,  avoiding any extensive matrix expansion. Consequently, the two-step approach not only requires significantly more computational time but also demonstrates a higher rate of increase in time complexity. Fig.~\ref{fig:fig2}d presents the average observer order and its standard deviation. The data-driven approach achieves a lower average order and a reduced standard deviation simultaneously. This is perhaps because errors introduced during the system identification stage can propagate into the subsequent observer augmentation stage, thereby introducing uncertainties into the entire two-stage design. This error propagation necessitates, on average, higher-order observers with greater variability in their design.
    	\begin{figure} \label{fig:fig2}
		\centering	\includegraphics[width=1\linewidth]{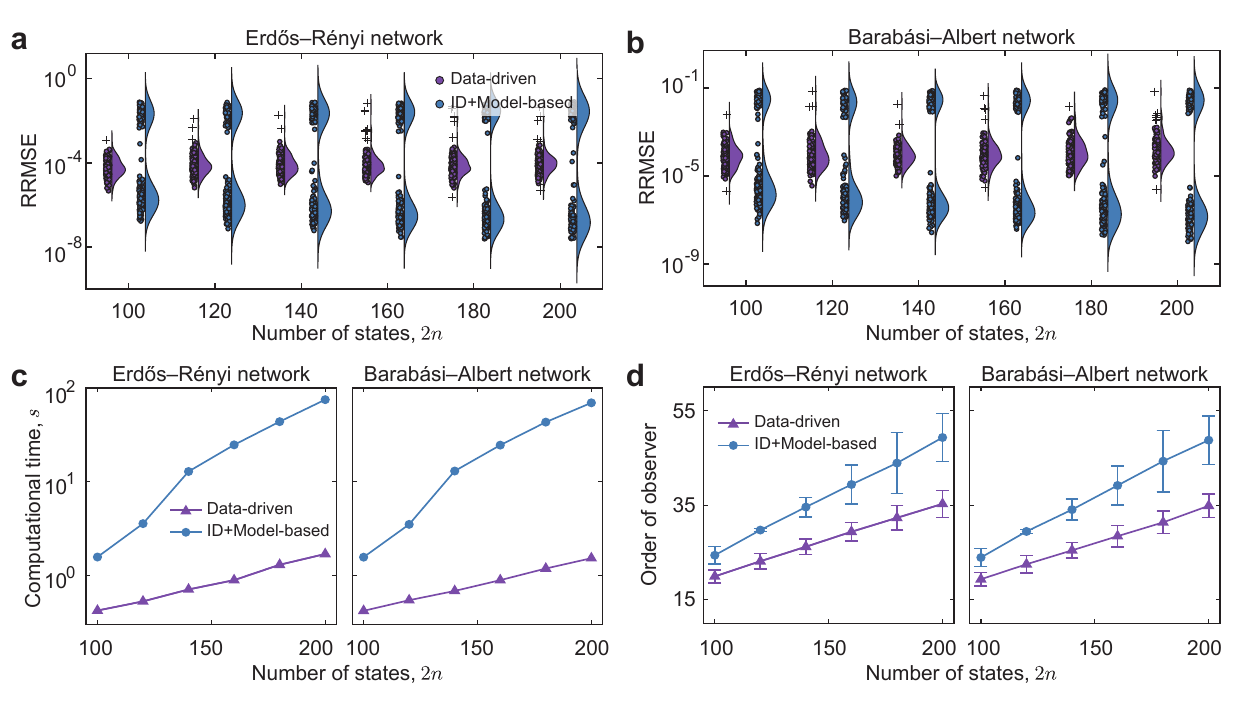}
        \caption{ { 
        {\textbf{Performance comparison between data-driven and identification-and-model-based functional observer designs.}} \textbf{a,b} RRMSE distribution for data-driven (purple) and ID+model-based (blue) observers in Erdős--Rényi (\textbf{a}) and Barabási--Albert (\textbf{b}) networks. Each data point represents an average over 100 independent network realizations.
\textbf{c} Computational time required for observer design in ER (left) and BA (right) networks across varying system dimensions \(2n\).
\textbf{d} Average observer order and standard deviation for both design methods.
Benchmarks were performed on randomly generated ER and BA networks with \(2n \in [100,200]\) states (each node is a 2-dimensional subsystem; see Methods). Systems have \(m = \lfloor 2n/5 \rfloor\) inputs, \(p = \lfloor n/2 \rfloor\) outputs, and \(r = \lfloor n/10 \rfloor\) functional states, with dedicated node assignments, where $\lfloor a \rfloor$ takes the floor of $a$. All networks have average degree \(\langle k\rangle=10\) and BA networks have the initial number of nodes $m_0=20$. Systems are stabilized by normalizing the state matrix by \(1.01\rho(\mathbf{A})\), where $\rho(\mathbf A)$ denotes the spectral radius of $\mathbf A$. The data-driven functional observer was trained directly on historical input-output data of length \(T = T_{\rm IO}^* + 100\) (sufficient for system identification), while the model-based approach first identified system matrices using the MOESP algorithm (see ``Methods'') with
true state dimensions given, followed by conventional functional observer design\cite{fernando2010functional}. To test the observers' performance, the initial states of the original systems are sampled from the uniform distribution on $(-1,1)$ while all observers start from zero. RRMSE is computed over the time period \([T,2T]\) with step input $u_i(t)=1, \forall t>0,i=1,2,...,m$.  } }
		\label{fig:fig2}
	\end{figure}

 \noindent \textbf{Network structure and observer performances.}  When the available historical data length is relatively small, the observer design procedure must incorporate the extended functional state data $\mathbf W$.   In this case, collecting full-dimensional state data $\mathbf{X}$ is either unaffordable or intractable, especially for large-scale networks, while only a part of state information is accessible.  Given partial information $\mathbf{W}=\mathbf{P}\mathbf{X}$, where matrix $\mathbf{P}$ consists of sub-rows of the identity matrix to index the available partial state data, it is still possible to recover functional states with satisfactory performance in our framework. 
   By directly substituting historical data $\mathbf{U}$, $\mathbf{Y}$, $\mathbf{Z}$, and $\mathbf{W}=\mathbf{P}\mathbf{X}$ into Algorithm 1 in ``Methods'', we derive the matrix $\mathbf{R}_s$ under partial information. 
    However, insufficient partial information may prevent the iterative procedure from guaranteeing satisfaction of Eqs. (\ref{condnew1}) and \eqref{condnew1-add}, indicating that the dynamics of the functional state cannot be fully recovered from the given partial information. 

    In Fig. \ref{fig:fig3}, we explore how network structure and data availability influence observer performance. 
Fig.~\ref{fig:fig3}a displays the RRMSE as a function of the percentage of partial information for both ER and BA random networks. Here,  partial information percentage is defined as $({\rm row}(\mathbf W) + {\rm row}(\mathbf Z))/(2n)$, where ${\rm row}(\mathbf W)$ and ${\rm row}(\mathbf Z)$ denote the number of rows in $\mathbf{W}$ and $\mathbf{Z}$, respectively, and $2n$ is the full state dimension. We find that higher partial information percentages generally yield lower RRMSE, indicating improved functional state recovery.  Notably, at 100\% partial information, the extended-state design achieves lower RRMSE than the IO-data based design for both network types, likely because the latter relies on computed projections ${\mathbf X}^{\rm proj}$ susceptible to rounding errors, while the former uses the precise $\mathbf{X}$.
  We classify an observer as convergent if its RRMSE is below a threshold $\epsilon=8\%$, chosen to exclude divergent cases and select observers that capture the majority of the functional state dynamics. Fig.~\ref{fig:fig3}c shows that the percentage of convergent realizations remains at zero when partial information is insufficient, but increases rapidly once a network-dependent threshold is exceeded. For example, ER networks with $p_{\text{edge}}=0.2$ require about $50\%$ partial information, while BA networks with average degree $\langle k \rangle=20$ require approximately $60\%$. Denser networks exhibit lower thresholds, likely because increased connectivity propagates state information more effectively. Notably, convergence percentages decline sharply when partial information exceeds approximately $90\%$. Excessive information may introduce irrelevant data, complicating the recovery of target dynamics. Additionally, ER networks achieve higher convergence percentages than BA networks, which we attribute to the uniform sampling of partial information in ER networks versus the need to sample critical hub nodes (nodes with high degrees) in scale-free BA networks. Fig.~\ref{fig:fig3}b presents the average observer order versus the percentage of partial information. The observer order increases approximately linearly with the percentage of partial information below the convergence threshold. These findings underscore the importance of selecting an appropriate amount of partial information for effective observer design.

	\begin{figure} \label{fig:fig3}
		\centering
	       \includegraphics[width=1\linewidth]{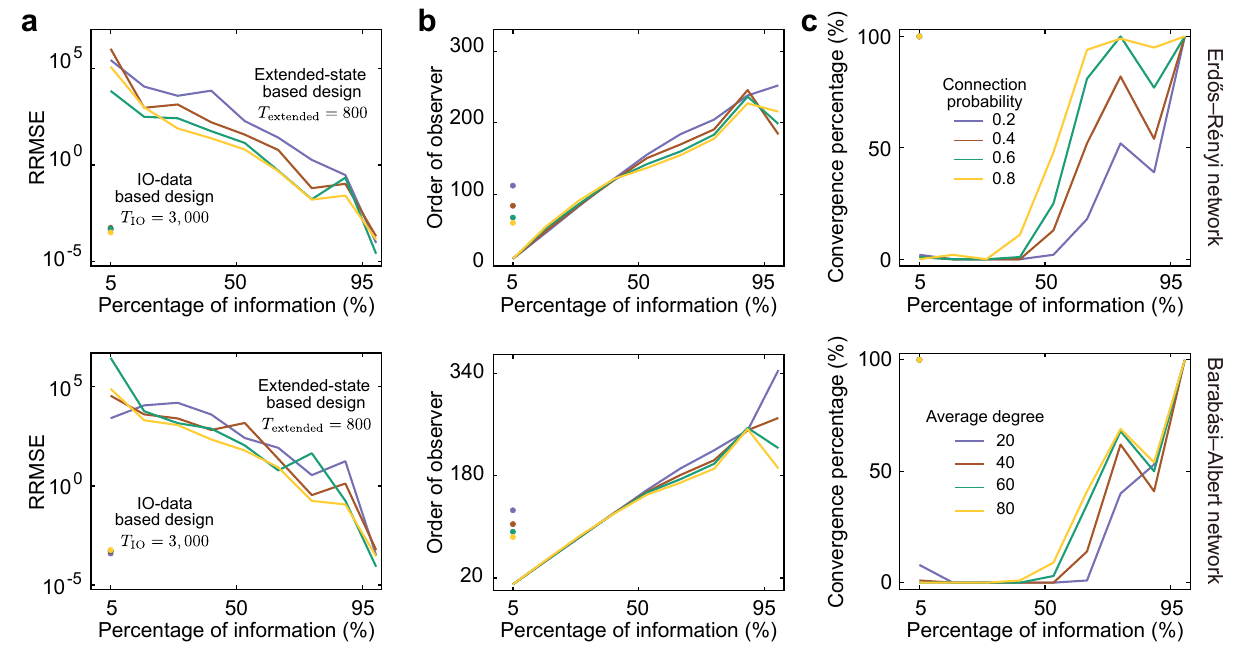}
		\caption{ {\textbf{Observer performance with partial state information.}} \textbf{a} RRMSE as a function of the percentage of partial information.
\textbf{b} Average observer orders across different information levels.
\textbf{c} Percentage of convergent realizations as a function of the percentage of partial information. Lines correspond to the extended-state based design with data length $T=800$, while dots correspond to the IO-data based design with a much larger data length $T=3000$. 
  We generate ER and BA random networks using the same settings as Fig.~\ref{fig:fig2} with fixed parameters $2n=500$, $m=80$, $p=100$, $r=10$, and varying parameters $p_{\text{edge}}$ for ER networks and average degrees $\langle k \rangle$ for BA networks, but only select the observable network realizations to ensure that all partial state information ${\mathbf P}\mathbf X$ corresponds to the observable component.
  {The rows of $\mathbf Z$ and $\mathbf W$ are uniformly sampled from rows of $\mathbf X$}.
  The RRMSE is taken over the period $\left[T,2T\right]$ with step input $u_i(t)=1, \forall t>0,i=1,2,...,m$. All results are averaged over 100 independent realizations.
        }
        
		\label{fig:fig3}
	\end{figure}

	 	\noindent \textbf{Extension to nonlinear systems with Koopman operators.}
	Our framework can be extended to nonlinear systems with the aid of Koopman operators. 
	Consider a discrete-time nonlinear system:
\begin{subequations}\label{eq:nonlinear_system_Koopman}
		\begin{align}
		\mathbf{x}({k+1}) &= f\left(\mathbf{x}(k), \mathbf{u}(k)\right), \\
		\mathbf{y}(k) &= g\left(\mathbf{x}(k)\right),
		\end{align}
	\end{subequations}
	where $f(\mathbf{x}(k),\mathbf{u}(k))$ and $g(\mathbf{x}(k))$ are nonlinear functions, $\mathbf x(k)\in {\mathbb R}^n$, and $\mathbf u(k)\in {\mathbb R}^m$. The Koopman operator theory\cite{koopman1931hamiltonian} provides a pathway to linearize such systems by lifting the state \( \mathbf{x}(k)\) into a higher-dimensional space using a set of lifting functions (also called observables) \( \phi_1, \dots, \phi_N \) with \( \phi_i: \mathbb{R}^n \to \mathbb{R} \) (see Supplementary Note 5). If there exists a finite collection of linearly independent observables such that the lifted state
$\Phi(\mathbf{x}(k)) = [\phi_1(\mathbf{x}(k)), \dots, \phi_N(\mathbf{x}(k))]^\intercal \in \mathbb{R}^N$
evolves linearly along all trajectories of~(\ref{eq:nonlinear_system_Koopman}), and the output \( \mathbf{y}(k) \) is a linear function of \( \Phi(\mathbf{x}(k)) \) and \( \mathbf{u}(k) \), then the system is said to admit a Koopman linear embedding\cite{shang2024willems}. In cases where an exact finite-dimensional linear embedding does not exist, a common practice is to approximate the dynamics using a finite set of observables, as employed in the extended dynamic mode decomposition\cite{williams2015data,lusch2018deep}.

	
	

    For nonlinear systems admitting an (approximate) Koopman linear embedding, the lifted state $\mathbf{\xi}(k) = \Phi\left(\mathbf{x}(k)\right) \in \mathbb{R}^{N}$ satisfies a linear equation as follows
\begin{subequations}\label{eq:linear_system_after_Koopman}
    \setlength\abovedisplayskip{6pt}
    \setlength\belowdisplayskip{3pt}
    \begin{align}
    \mathbf{\xi}({k+1}) &= \mathbf{A}_{\Phi}\mathbf{\xi}(k) + \mathbf{B}_{\Phi} \mathbf{u}(k), \\
    \mathbf{y}(k) &= \mathbf{C}_{\Phi}\mathbf{\xi}(k),
    \end{align}
\end{subequations}for some matrices $(\mathbf A_\Phi,\mathbf B_{\Phi},\mathbf C_{\Phi})$ of appropriate dimensions. 
Typically, $N \gg n$, and the matrix pairs $(\mathbf{A}_{\Phi},\mathbf{B}_{\Phi})$ and $(\mathbf{A}_{\Phi}, \mathbf{C}_{\Phi})$ in (\ref{eq:linear_system_after_Koopman}) may be uncontrollable or unobservable due to the introduction of auxiliary correlated lifting functions (see Supplementary Note 5 for an example). Consequently, conventional full-state Luenberger observers become infeasible due to the high dimension $N$ and potential unobservability of \eqref{eq:linear_system_after_Koopman}. Our proposed data-driven functional observer naturally addresses these limitations. Without loss of generality, we assume the original state $\mathbf{x}(k)$ constitutes part of the Koopman embedding state, such that $\mathbf{\xi}(k)=[\mathbf{x}^\intercal(k),\mathbf{\zeta}^\intercal(k)]^\intercal$ with $\mathbf{\zeta}(k)\in {\mathbb R}^{N-n}$. A linear function $\mathbf{z}(k)=\mathbf{F}{\mathbf x}(k)$ can then be expressed as:
\begin{equation}\label{koopman}
\mathbf{z}(k) = \mathbf{F}_\mathbf{{x}}\mathbf{\xi}(k),
\end{equation}
where $\mathbf{F}_\mathbf{{x}} \doteq  {\mathbf{F}}[\mathbf{I}_{n}, \, \mathbf{0}_{n\times (N-n)} ]$. The state-space model thus takes the form of \eqref{eq:linear_system_after_Koopman}--\eqref{koopman}. For such linear representations, our data-driven observer design estimates the functional state $\mathbf{z}(k)$ without identifying $(\mathbf{A}_{\Phi},\mathbf{B}_{\Phi},\mathbf{C}_{\Phi})$. This approach contrasts with traditional methods that first determine $(\mathbf{A}_{\Phi},\mathbf{B}_{\Phi},\mathbf{C}_{\Phi})$ via extended dynamic mode decomposition before implementing model-based Luenberger observers\cite{surana2016linear}. By circumventing this process and focusing directly on $\mathbf{z}(k)$ rather than all lifted variables, our method achieves substantially lower observer orders and improved computational efficiency. 
Notably, for nonlinear systems with exact finite-dimensional Koopman embeddings, our IO-data based design operates without explicit lifting functions (since no extended state data is needed) when sufficient historical data is available.

 	\noindent \textbf{Observer design with noisy data.}
	In practical scenarios, historical data are often contaminated by noise. In this case, the data matrix $\mathbf{D}= \left[{\mathbf{U}_p}^{\intercal}, {\mathbf{Y}_p}^{\intercal}, {\mathbf{Y}_f}^{\intercal}, {\mathbf{Z}_p}^{\intercal}\right]^{\intercal}$ may become full row rank, thereby restricting the existence of a stable functional observer. To address this issue, noise mitigation strategies can be employed. When some prior knowledge about the noise characteristics is available, such information can be leveraged to obtain more accurate estimator parameters\cite{verhaegen2007filtering,baggio2021data}. A particularly relevant case arises when the historical data are affected by additive noise:
	\begin{equation} \label{noise-set}
	\begin{array}{l}
	\mathbf{\bar{Y}}_{0,T} = \mathbf{Y}_{0,T} + \mathbf{V}_{0,T}, \\
	\mathbf{\bar{Z}}_{0,T} = \mathbf{Z}_{0,T} + \mathbf{J}_{0,T}, 
	\end{array}
	\end{equation}
	where $\mathbf{V}_{0,T}$ and $\mathbf{J}_{0,T}$ are random matrices with i.i.d. zero-mean entries of variances $\sigma^2_\mathbf{Y}$ and $\sigma^2_\mathbf{{Z}}$, respectively, $\mathbf{V}_{0,T}$ and $\mathbf{J}_{0,T}$ are unrelated with the truth value $\mathbf{Y}_{0,T}$ and $\mathbf{Z}_{0,T}$.  Partition $\mathbf{\bar Y}_{0,T}$ and $\mathbf{\bar Z}_{0,T}$ into past and future components: $\mathbf{\bar{Y}}_p = \mathbf{\bar{Y}}_{0,T-1}$, $\mathbf{\bar{Y}}_f = \mathbf{\bar{Y}}_{1,T-1}$, $\mathbf{\bar{Z}}_p = \mathbf{\bar{Z}}_{0,T-1}$, and $\mathbf{\bar{Z}}_f = \mathbf{\bar{Z}}_{1,T-1}$.
	
	To account for noise, we modify the estimation of the observer parameters based on compensation for pseudoinverse operations and truncated singular value decomposition (SVD). Here, we take the parameters in \eqref{general-solution} for a Darouach observer as an example. Specifically, denoting $\mathbf{\bar D
	}\doteq \left[{\mathbf{U}}_p^{\intercal},\bar {\mathbf{Y}}_p^{\intercal}, \bar {\mathbf{Y}}_f^{\intercal}, \bar {\mathbf{Z}}_p^{\intercal}\right]^{\intercal}$, an asymptotically unbiased estimation of ${\mathbf{Z}_f}\mathbf{D}^\dagger$ in \eqref{general-solution} reads as (Supplementary Note~6) 
	\begin{equation}\label{eq:noisy_estimation}
	\frac{1}{T-1} \bar {\mathbf{Z}}_f\mathbf{\bar{D}}^{\intercal}\left(\frac{1}{T-1} \mathbf{\bar{D}} \mathbf{\bar{D}}^{\intercal}-\mathbf{\Omega}\right)^{ \dagger}
	\end{equation}
	where
	\[
	\mathbf{\Omega} = \begin{bmatrix}
	0 & 0 & 0 & 0\\[1mm]
	0 & \sigma^2_\mathbf{Y} I_p & 0 & 0\\[1mm]
	0 & 0 & \sigma^2_\mathbf{Y} I_p & 0\\[1mm]
	0 & 0 & 0 & \sigma_\mathbf{Z}^2I_r
	\end{bmatrix}.
	\] Furthermore, denoting $\lambda_1 \geq \lambda_2 \geq \cdots \geq \lambda_{m+2p+r} \geq 0$ as the singular values of $\frac{1}{T-1} \mathbf{\bar{D}} \mathbf{\bar{D}}^\intercal$, an approximation $\tilde k_\mathbf{D}$ of $k_\mathbf{D}\doteq {\rm rank}(\mathbf{D})$ can be $\tilde k_\mathbf{D} = \left| \left\{ \lambda_i : \lambda_i \geq \eta \max\{\sigma_{\mathbf{Y}}^2,\sigma_{\mathbf{Z}}^2\} \right\} \right|$,
	where $\eta$ is a number to separate the dominant singular values from the system and the spurious nonzero singular values from the noise (when the signal-to-noise ratio is large, $\eta$ can be some number near one). For the SVD $ \frac{1}{T-1} \mathbf{\bar{D}}\mathbf{\bar{D}}^{\intercal}-\mathbf{\Omega}=\left[\begin{array}{ll}
	\mathbf{U}_1 & \mathbf{U}_2
	\end{array}\right] \mathbf{S}\left[\begin{array}{l}
	\mathbf{V}_1 \\
	\mathbf{V}_2
	\end{array}\right]$, where $\mathbf{U}_1$ is the first $\tilde k_\mathbf{D}$ columns of $\left[\mathbf{U}_1~\mathbf{U}_2\right]$, the row space of $\mathbf{U}_2^\intercal$ serves as a good estimation of the left null space of $\mathbf{D}$. 

    In scenarios where the noise variances are unknown, we can employ an SVD-based strategy to robustly identify the dominant data subspace from noisy data. The key is to compute the SVD of the data matrix and use the pronounced gap in its singular value spectrum to separate the signal subspace from the noise subspace. By truncating the SVD to retain only the dominant components, we obtain a denoised approximation of the data matrix $\mathbf D$, which is subsequently used to design functional observers. A detailed description of the methodology, along with numerical validation demonstrating its performance across various noise levels, is provided in the Supplementary Note~6 and Figure S5.

\begin{figure}[htbp]
		\centering
\includegraphics[width=1\linewidth]{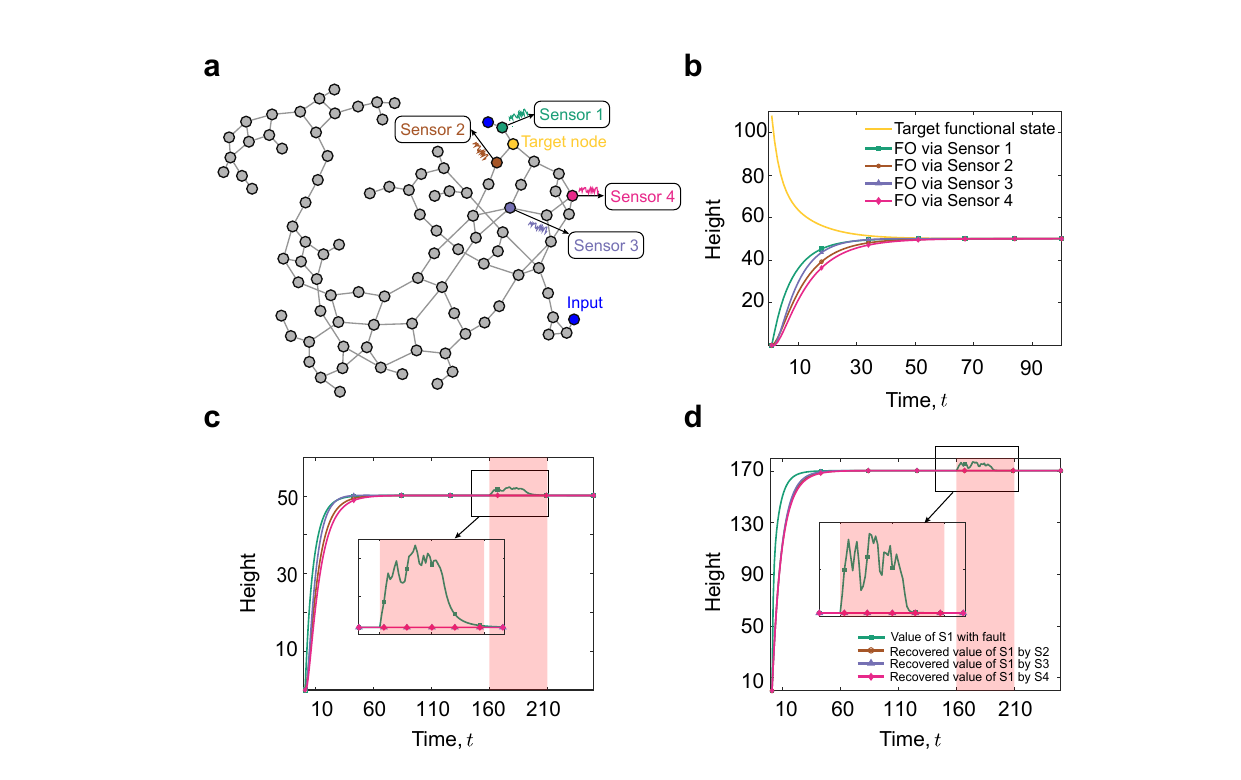}
        \caption{{\small 
        {\textbf{Sensor fault detection and recovery in a water network.}} \textbf{a} The structure of EPANET's network 3. Each edge is bidirectional (except the edges from input nodes to state nodes) and each state node has a self-loop which is omitted. The reservoirs whose water heads can be arbitrarily set by the controller serve as the system inputs (blue nodes), while the remaining 95 nodes evolve according to the network dynamics.  \textbf{b} The trajectory of the target node's height (true functional state) and its estimations from four independent functional observers (FO), each using measurements of one normal sensor. \textbf{c} The trajectories of the aforementioned four functional observers, where sensor 1 is under attack.  Red areas indicate the time intervals of sensor attack. \textbf{d} Estimations of the value of sensor 1 from four independent functional observers, where each uses measurement of one normal sensor to estimate the value of sensor 1. In our numerical experiments, we simulate the transport of engine oil at \(10^\circ\)C, with \(\rho = 885\,\mathrm{kg/m^3}\), and \(\mu = 582.95\,\mathrm{mPa\cdot s}\). Pipe diameters $D_{ij}$ are sampled uniformly from 10 to 20~m and lengths $L_{ij}$ from 3000 to 4000~m. Continuous-time dynamics are discretized using a forward-Euler scheme with a time step of $\Delta t = 0.01$~s. The resulting system matrix is normalized by its spectral radius to ensure numerical stability. The historical data are generated with Gaussian random inputs. }}
        \label{fig:water}
        \end{figure}

	\noindent \textbf{Sensor fault detection and recovery in water networks.} To validate the practical applicability of our data-driven functional observer framework, we consider the problem of sensor-fault detection in an EPANET 3 water-supply network\cite{rossman1994epanet}. Following the modeling approaches in refs.\cite{burgschweiger2009optimization, boulos2006comprehensive}, the network is represented as a directed graph comprising 97 nodes--specifically, 2 reservoirs, 3 tanks, and 92 junctions, shown in Fig.~\ref{fig:water}a. 	At each junction or tank $i$, the head dynamics obey the mass-balance equation
		$\dot h_i(t) = \sum_{j\in\partial i} Q_{ji},$
		where $h_i$ is the hydraulic head at node $i$, $Q_{ji}$ denotes the flow rate from node $j$ into node $i$, and $\partial i$ is the neighbor set of node $i$. Under laminar flow conditions, the Darcy-Weisbach relation reduces to a linear head-loss model\cite{brown2002history}:
		\[
		h_i - h_j = \frac{128\,\mu\,L_{ij}}{\rho\,\pi\,g\,D_{ij}^4} Q_{ij},
		\]
		with $\mu$ the dynamic viscosity, $\rho$ the fluid density, $L_{ij}$ and $D_{ij}$ the length and hydraulic diameter of the pipe between nodes $i$ and $j$, respectively, and $g$ the acceleration due to gravity.

			The basic idea of sensor-fault detection by functional observers is based on cross-validation: assuming that the majority of sensor measurements remain normal, any subset of normal sensors drawn from the intact pool will yield consistent estimates of a given functional state (provided it is functionally observable). Under stochastic sensing, if all subsets produce statistically similar estimates, no sensor anomaly is present. Otherwise, divergence among estimates indicates the presence of a compromised sensor subset, enabling rough localization of the fault via cross-validation. Once the sensors with fault are located, one can treat their values as new functional states, then construct functional observers using the output of other sensors to recover the true values of these sensors.  Consider a simple yet insightful example, where four pressure sensors are deployed at strategically chosen junctions (see Fig.~\ref{fig:water}a). Our target functional state is the head at the critical junction highlighted in yellow if Fig.~\ref{fig:water}a. We design four independent functional observers, each using measurements from one sensor only. Under nominal (fault-free) conditions, all observers accurately track the functional state, with steady-state trajectories shown in Fig.~\ref{fig:water}b. To evaluate fault detection and recovery, we introduce an additive bias on one sensor (green node in Fig.~\ref{fig:water}a, namely, sensor 1) during the interval $t\in[160,190]$~s:
	$\mathbf{y}(k+1) = \mathbf{C}\,\mathbf{h}(k) + \mathbf{\mu}(k)$, where $\mathbf{\mu}(k)$ is a random offset sampled from the uniform distribution on the interval $(0,10)$. Fig.~\ref{fig:water}c plots the observer estimates during the attack, and highlights the divergence of the compromised observer compared to the unaffected ones. This discrepancy enables automatic isolation of the faulty sensor: only the observer relying on the attacked measurement fails to reconstruct the target state, demonstrating the efficacy of our data-driven scheme in promptly detecting sensor anomalies.  
    We next treat the value of Sensor 1 as the functional state and construct four independent functional observers using the output of Sensor 2,3, and 4, respectively. As shown in Fig.~\ref{fig:water}d, all three functional observers successfully recover the true value of Sensor 1. Although we only present the scenario where only one sensor is under attack, it is not hard to extend it to the general case where a subset of sensors are under attack as in ref.\cite{fawzi2014secure}.

  \noindent \textbf{Learning feedback control laws using functional observers in load frequency control.}   With the continuous development of power systems, traditional single-area networks have evolved into interconnected multi-area systems. While interconnected power systems (IPS) enhance power supply reliability and stability, they also introduce greater operational complexity and challenges. Load disturbances can cause disruptive effects, such as deviations of frequency and tie-line power from their desired values\cite{pan2002adaptive}, posing significant threats to the economic and safe operation of IPS.  Load frequency control (LFC) is a key mechanism in power grids for balancing load consumption with power generation. It is crucial for preventing frequency deviations in large-scale multi-area IPS from damaging equipment, impairing power quality, or triggering large-scale blackouts. LFC in multi-area power systems requires real-time estimation of control-relevant quantities such as area control errors (ACE) and integrated frequency deviations, which are typically linear combinations of system states rather than individual state variables. These functional states are often not directly measurable and must be reconstructed from available sensor data across different control areas. By designing a data-driven functional observer-based LFC feedback control strategy, we can maintain frequency deviations and tie-line power deviations within acceptable ranges when the system is subject to load disturbances, thereby ensuring the stability of the power system and further demonstrating the versatility of our framework.  

\begin{figure}[H]
    \centering
 	\includegraphics[width=1\linewidth]{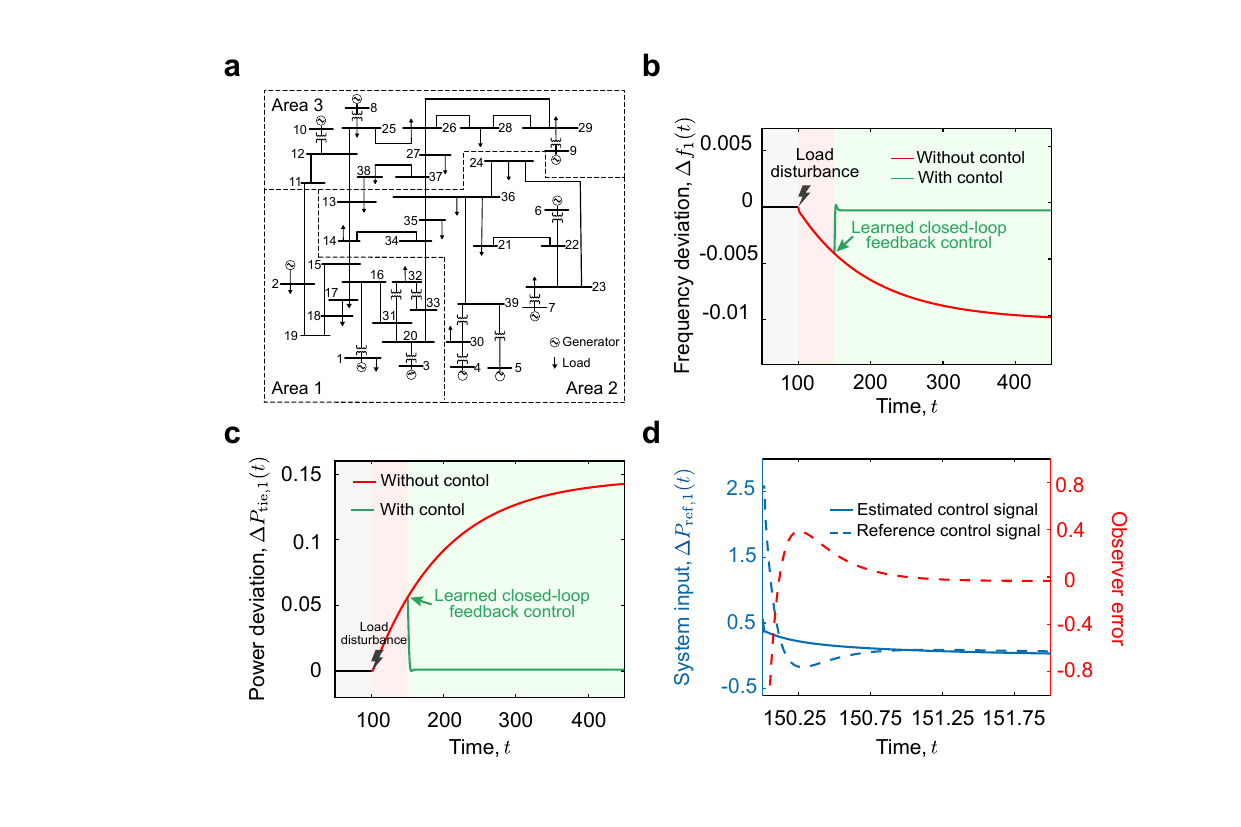}
    \caption{\textbf{a} Partition of the IEEE 39-bus system, with Area 1 in the bottom-left, Area 2 in the bottom-right, and Area 3 in the top. \textbf{b} The open-loop and closed-loop frequency deviations (green region, after $t=150s$) of Area 1 when a persistent $0.1$  step load disturbance is imposed on Area 1 at time $t=100s$ (red region). \textbf{c} The open-loop and closed-loop tie-line deviations of Area 1. \textbf{d} Control input and the error of learning a control law using functional observers after $t=150s$.  } 
    \label{fig:LFCieee39}
\end{figure}

   Here, we utilize the IEEE-39 New England power grid as a case study. This system consists of 39 nodes (29 load nodes and 10 generator nodes), which are divided into three area ($N=3$) interconnected by tie-lines\cite{shangguan2021adjustable}, as shown in Fig.~\ref{fig:LFCieee39}a. As in refs.\cite{shangguan2021adjustable,ma2014distributed}, the dynamics of the frequency deviation $\Delta f_i$ and tie-line power deviation $\Delta P_{{\rm tie},i}$ of each area $i$ follows a linear equation, with the augmented multi-area LFC state space equation given in ``Mthods''. In particular, each generator is equipped with a dedicated sensor. Let $x(t)$ be the corresponding state vector. 
To ensure system stability and meet closed-loop control performance requirements, a global stabilizing state feedback control law of the form $\mathbf{u}(t)=\mathbf{L}{\mathbf{x}}(t)$ can be designed to achieve the prescribed closed-loop system performance. The gain matrix $\mathbf{L}$ can be derived through data-driven feedback controller design using the method presented in ref.\cite{de2019formulas}.
		However, a critical limitation of this control law lies in the unavailability of state variables for feedback control. Rather than implementing a Luenberger state observer to estimate ${\mathbf{x}}(t)$, we adopt a reduced-order functional observer that directly estimates the functional state $\mathbf z(t)=\mathbf{L}{\mathbf{x}}(t)$, offering both implementation advantages and computational efficiency.
		In particular, we use the open-loop data generated by random inputs and random initial states to learn a functional observer for estimating ${\mathbf{z}}(t)$ and use the estimate as control input. Due to the separation principle in the design of observer-based feedback controllers\cite{trinh2011functional,montanari2025duality}, the stability of the closed-loop system is guaranteed if the functional observer and the direct state feedback system are both stable.
       
       In Fig.~\ref{fig:LFCieee39}b and Fig.~\ref{fig:LFCieee39}c, during the initial steady-state operation phase, the frequency deviation of area 1 and the tie-line power deviations between areas remain in a relatively stable equilibrium state (blue region). However, after \(t = 100s\) (red region), when a $0.1$ step load disturbance is applied to area 1, the frequency deviations and tie-line power deviations deviated from this equilibrium state, and significant changes occurred without control, which reflects the impact of load disturbances on system dynamics. Fig.~\ref{fig:LFCieee39}d shows the error and tracking situations when learning the control law through the functional observer. The converging error curve demonstrates the observer's ability to track the control law over time, with the error gradually approaching zero, indicating that the learning accuracy is improved. After applying the learned closed-loop feedback control (green region in Fig.~\ref{fig:LFCieee39}b and  Fig.~\ref{fig:LFCieee39}c, after \(t = 150s\)), the frequency deviation and tie-line power deviation of area $1$ are significantly reduced (blue line). The control strategy adjusts the power output according to the observed system states (such as phase and frequency errors), thereby restoring and maintaining system stability. 

        

\begin{figure}[H]
    \centering
 	\centering		\includegraphics[width=1.0\linewidth]{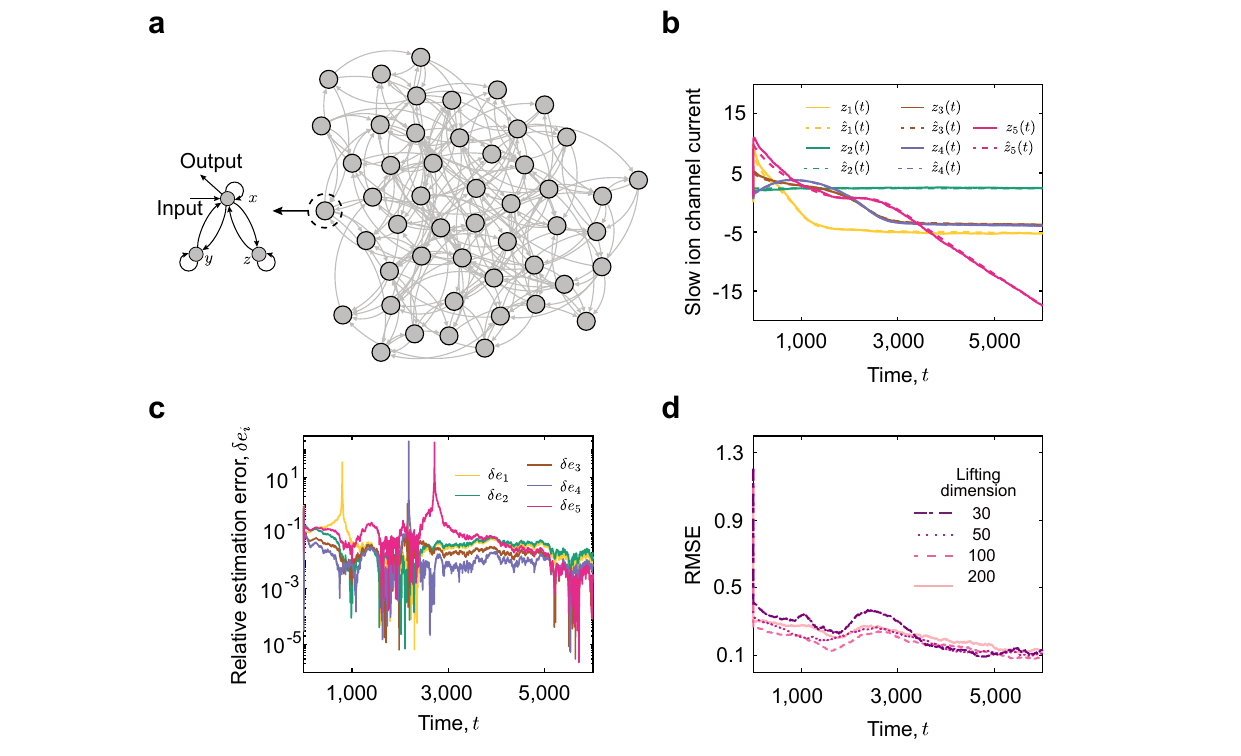}
  \caption{\textbf{Target estimation of nonlinear neuron networks using limited sensors.} \textbf{a} The interaction network of a neuron system with $N=50$ neurons. Each node contains 3 states $x,y$, and $z$, among which $80$ sensor locations and $5$ target states are randomly selected. \textbf{b} Representative trajectories of the five target states (solid lines) and their estimates (dashed lines) obtained using the data-driven functional observer with $n_z=50$ thin plate spline radial basis functions. \textbf{c} Corresponding relative estimation errors $\delta e_i = |\hat{x}_i - x_i|/|x_i|$ for the trial shown in \textbf{b}.  \textbf{d} Time-dependent RMSE ($R_t$) for different lifting dimensions (30, 50, 100, and 200), with each data point representing the average over 100 independent experimental trials. The system parameters are set to $a=1,b=3, c=1, d=5, \gamma=0.001, s=1, x_R=-1.6$, and $\epsilon=N$.  }
		\label{fig:koopmanNeuronAB}
\end{figure}

	 \noindent \textbf{Target estimation of nonlinear neuron networks using limited sensors.} To further illustrate the application of our approach for nonlinear systems, we apply the data-driven observer framework to target estimation in neuron networks modeled by the classic Hindmarsh-Rose system\cite{hindmarsh1984model}. This model captures the  spiking-bursting  behavior of the $i$-th neuron, given by
\begin{equation}
\begin{aligned}
& \dot{x}_i=y_i-a x_i^3+b x_i^2-z_i+\frac{\epsilon}{N} \sum_{j \in \partial_i^{+}}\left(x_j-x_i\right)+I_i, \\
& \dot{y}_i=c-d x_i^2-y_i, \\
& \dot{z}_i=\gamma\left(s\left(x_i-x_R\right)-z_i\right),
\end{aligned}
\end{equation}
where $x_i(t)$ is the membrane potential, $y_i(t)$ describes fast ion channel dynamics, and $z_i(t)$ represents slow ion channel dynamics. The external current $I_i$ acts as an input for neuron $i$ and $\partial_i^{+}$ denotes the set of neurons to which neuron $i$ is connected, and $a,b,
c,d,\gamma,s$ are constants. In our simulation, $N=50$ neurons are coupled over a network of $50$ nodes sampled from ER random networks with connectivity probability $p_{\rm edge} = 0.1$, whose topology is given in  Fig.~ \ref{fig:koopmanNeuronAB}a.
The time step is set to $T=6000$ to investigate long-term dynamic characteristics. Among the $50$ neurons (the dimension of the total states is $150$, since each neuron has $3$ state variables), we place $80$ sensors to measure different state variables, and $5$ states are chosen as target functional states.  The historical input-output data are generated by imposing random inputs and initial states. 
The Koopman linear embedding method is applied to lift the nonlinear model into a linear space using lifting functions $\phi_i(\mathbf{x})$ that include the state variables themselves  and
$n_z$ thin plate spline radial basis functions\cite{korda2018linear} $\phi_i(\mathbf{x}) = \left\|\mathbf{x} - \mathbf{x}_{0i}\right\|_2^2 \log \left( \left\|\mathbf{x} - \mathbf{x}_{0i}\right\|_2^2 \right)$ with centers $\mathbf x_{0i}$ uniformly distributed in the unit box.  Fig.~\ref{fig:koopmanNeuronAB}b collects all estimated target states from the data-driven functional observer via $n_z=50$ and the corresponding true system states, with the estimation errors given in Fig.~\ref{fig:koopmanNeuronAB}c. The results show that the estimated trajectories closely align with ground-truth trajectories, demonstrating the effectiveness of the proposed data-driven observers even for nonlinear systems. 





To evaluate the impact of the Koopman operator's lifting dimension on the observation performance, some comparative experiments were conducted (with lifting dimensions set to $30, 50, 100$ and $200$). To be specific, we conduct $N_r=100$ experimental trials, each initialized by randomly generating an initial state ${x}_0^i$. Denote the sequence $\left\{{x}_0^i, {x}_1^i, \ldots, {x}_{T_0}^i\right\}$ as the true state sequence and $\left\{\hat{{x}}_0^i, \hat{{x}}_1^i, \ldots, \hat{{x}}_{T_0}^i\right\}$  the estimated one, with $T_0$ the time horizon of simulation. The estimation accuracy is evaluated using the time-dependent Root Mean Square Error (RMSE): $ R_t=\sqrt{\frac{1}{N_r} \sum_{i=1}^{N_r}\left\|{x}_t^i-\hat{x}_t^i\right\|^2}$, $t=1, \ldots, T_0$. The RMSE curves in Fig.~\ref{fig:koopmanNeuronAB}d indicate that increasing the Koopman lifting dimension improves the performance of the proposed observer before $n_z$ reaching $100$.
However, excessively large $n_z$ is not optimal, as seen from the case with $n_z=200$. This is perhaps because higher-dimensional basis functions may induce overfitting so that the designed observer may perfectly fit training data but accumulate rapid errors on test data. More examples of our method in target state estimation of empirical nonlinear network systems can be found in Supplementary Note 7 and Figures S6 and S7.  

	\section*{Discussion}

State estimation is crucial for monitoring, supervision, and control of dynamic systems. However, reconstructing the complete set of state variables is often infeasible due to limited sensing, high dimensionality, or unobservable dynamics. Traditional observer design relies on explicit system models, which can fail when these models are unavailable or inaccurate. Motivated by successful data-driven approaches in physics and engineering\cite{korda2018linear,Brunton2016SINDy,Brunton2016SINDy,yabe2022toward}, we propose a unified framework with theoretical guarantees for designing functional observers directly from data, enabling real-time target estimation without system identification.

Our results introduce a data-driven functional observability criterion based on Hankel matrices derived from historical trajectories. This criterion requires only input-output data and partial state measurements, making it applicable even to unobservable systems, and provides a tool for data-driven observability analysis and sensor placement. Beyond linear dynamics, the criterion suggests broader implications for nonlinear systems. It provides a data-based solution to the nonlinear subspace reconstruction problem discussed in ref.\cite{montanari2022functional-nonlinear}, that is, whether a subspace of a nonlinear system can be uniquely inferred from input-output sequences.  We also present methods for designing functional observers that match model-based designs in accuracy while excelling in partially observable regimes.

Compared with classical Luenberger-type observers, our approach reduces observer order when full state estimation is unnecessary, improving computational efficiency and enabling real-time implementation in large networks. Simulations and empirical case studies across diverse scenarios demonstrate the versatility of data-driven functional state estimation. As observers are algorithms that do not interfere with system dynamics, our approach is hardware-friendly, offering a flexible tool for sensor-rich environments like autonomous driving\cite{bertipaglia2024unscented}, where sensors can be allocated as target states and measurements. This opens new possibilities for fault detection, recovery, and sensor deployment. Our results suggest that functional observers are better suited for nonlinear systems, particularly when approximate linear embeddings exist.

Despite these advances, challenges remain. Our design relies on a single input-functional state-output trajectory satisfying certain persistent excitation conditions. Extending it to multiple trajectories would be promising, especially for nonlinear systems where a single trajectory may not capture the systems' global behaviors.   In the nonlinear setting, high-dimensional lifting bases improve representation but may increase computational load and overfit unseen data. When an exact finite linear embedding does not exist, developing methods to select lifting functions and quantify approximation errors is a demanding but challenging direction. Integrating functional observer design with feedback control in a fully data-driven framework is another open problem. This differs from data-enabled predictive control\cite{morato2024data}, which does not utilize internal states. Combining functional observers with emerging target controllers\cite{fernando2025existence,montanari2025duality} for large-scale systems is also an interesting future direction.
While our framework accounts for white noise, improving robustness against nonstationary or unknown noise distributions remains a challenge. Extending the methodology to data-based Kalman filtering for functional states is an appealing next step. A limitation of the current work is the reliance on sufficiently long {\emph{offline}} trajectories. While dynamic updates can handle time-varying networks, rigorous guarantees are lacking. Future work will explore adaptive online implementations, leveraging streaming data and machine learning techniques to maintain performance in evolving networks.

	\section*{Methods}\label{sec11}
\textbf{Model-based observer design.}
    For the system described in~\eqref{eq:system}, the Darouach observer takes the following form
\begin{equation*}
\left\{\begin{array}{l}
\mathbf{r}(t+1)=\mathbf{\bar{A}} \mathbf{r}(t)+\mathbf{\bar{B}}^{\mathbf{u}} \mathbf{u}(t)+\mathbf{\bar{B}}^{\mathbf y} \mathbf{y}(t), \\
\mathbf{\hat{z}}(t)=\mathbf{r}(t)+\mathbf{\bar{D} }\mathbf{y}(t),
\end{array}\right.
\end{equation*}where $\mathbf r(t),\hat {\mathbf z}(t)\in {\mathbb R}^r$ and  $(\mathbf{\bar{A}},\mathbf{\bar B^u}, \mathbf{\bar B^y}, \mathbf{\bar D})$ are parameters to be designed. 
Define the estimation error as $\mathbf{e}(t)=\mathbf{z}(t)-\mathbf{\hat{z}}(t)$.  Its dynamics is derived as
\begin{equation*}
\begin{aligned}
 \mathbf{e}(t+1)&=\mathbf{F}(\mathbf{A}\mathbf{ x}(t)+\mathbf{B} \mathbf{u}(t))-(\mathbf{r}(t+1)+\mathbf{\bar{D}} \mathbf{y}(t+1)) \\
& =\mathbf{\bar{A}} \mathbf{e}(t)+\left(\mathbf{F} \mathbf{A}-\mathbf{\bar{A}} \mathbf{F}-\mathbf{\bar{B}^y} \mathbf{C}-\mathbf{\bar{D} C A}+\mathbf{\bar{A} \bar{D} {C}}\right) \mathbf{x}(t)+\left(\mathbf{F} \mathbf{B}-\mathbf{\bar{B}^u}-\mathbf{\bar{D} C B}\right) \mathbf{u}(t).
\end{aligned}
\end{equation*}
The Darouach observer exists such that the estimation error converges asymptotically to zero from any initial guess $ \mathbf{\hat z}(t)$ if the following conditions hold\cite{Darouach2023FunctionalDA}
$$
\left\{\begin{array}{l}
\mathbf{F} \mathbf{A}-\mathbf{\bar{A}} \mathbf{F}-\mathbf{\bar{B}^y} \mathbf{C}-\mathbf{\bar{D}} \mathbf{C} \mathbf{A}+\mathbf{\bar{A}} \mathbf{\bar{D}} \mathbf{{C}}=0 ,\\
\mathbf{F} \mathbf{B}-\mathbf{\bar{B}^u}-\mathbf{\bar{D} C B}=0, \\
\mathbf{\bar{A}} \,\, {\text{Schur stable}}.
\end{array}\right.
$$Under these conditions, the error dynamics simplifies to $\mathbf{e}(t+1)= \mathbf{\bar A}\mathbf{e}(t)$.
A numerical algorithm for determining the parameters $(\mathbf{\bar{A}},\mathbf{\bar B^u}, \mathbf{\bar B^y}, \mathbf{\bar D})$ can be found in ref.\cite{darouach2000existence}. If a Darouach observer does not exist for a given $\mathbf{F}$, one can 
augment the functional matrix by adding new row $\mathbf{R}$, constructing an augmented functional matrix $\mathbf{F^{\prime}}=[\mathbf{R}^\intercal,\mathbf{F}^\intercal]^\intercal$ such that a Darouach observer exists with the augmented functional state $\mathbf{F'}\mathbf{x}(t)$\cite{fernando2010functional}. The order of the observer is defined as the dimension of the vector $\mathbf{F'}\mathbf{x}(t)$. An iterative algorithm for determining a suitable $\mathbf{R}$ can be found in ref.\cite{fernando2010functional}. When $(\mathbf{A},\mathbf{C})$ is observable, a Luenberger observer of order $n$ can be designed as $\mathbf{\hat x}(t+1)=\mathbf{A}\mathbf{\hat x}(t)+\mathbf{B}\mathbf{\hat u}(t)+\mathbf{L}(\mathbf{y}(t)-\mathbf{\hat C}\mathbf{\hat x}(t))$, where $\mathbf{L}$ is a designed gain matrix such that $\mathbf{A}-\mathbf{L}\mathbf{C}$ is Schur stable (whose eigenvalues can be arbitrarily assigned).


\noindent	{{\textbf{Subspace-based system identification.}}
		For the input sequence $\mathbf{U}$ and output sequence $\mathbf{Y}$ generated by system (\ref{eq:system}), let $\mathbf{U}_{0,s,N}$ and $\mathbf{Y}_{0,s,N}$ denote the corresponding Hankel matrices of order $s$. Leveraging the RQ factorization for sufficiently large $N$, we obtain\cite{verahegen1992subspace}
		\begin{equation}\label{eq:id}
		\left[\begin{matrix}
		\mathbf{U}_{0,s,N}\\\mathbf{Y}_{0,s,N}
		\end{matrix}\right] = 
		\left[\begin{matrix}
		\mathbf{R}_{11} & \mathbf{0} & \mathbf{0}\\
		\mathbf{R}_{21} & \mathbf{R}_{22} & \mathbf{0}
		\end{matrix}\right]
		\left[\begin{matrix}
		\mathbf{Q}_1\\\mathbf{Q}_2\\\mathbf{Q}_3
		\end{matrix}\right],
		\end{equation}where ${\mathbf{R}}_{11}\in {\mathbb R}^{sm\times sm}$ and ${\mathbf{R}}_{22}\in {\mathbb R}^{sp\times sp}$.
		Recall the $s$-step observability matrix is  $\mathbf{O}_{s}(\mathbf A, \mathbf C)= [\mathbf{C}^\intercal\; (\mathbf{CA})^\intercal \;\cdots \; (\mathbf{CA}^{s-1})^\intercal]^\intercal$, $s\in {\mathbb N}$. If the input sequence $U$ satisfies ${\rm rank}([\mathbf{X}_{0,N}^\intercal,\mathbf{U}_{0,s,N}^\intercal]^\intercal)=n+sm$ and $s\ge l(\mathbf{A},\mathbf{C})$,
		then $\text{span}_{\rm col}(\mathbf{O}_{s}(\mathbf A, \mathbf C)) = \text{span}_{\rm col}(\mathbf{R}_{22})$\cite[Theorem 9.1]{verhaegen2007filtering}, where $\text{span}_{\rm col}(\mathbf{M})$ takes the column space of matrix $\mathbf M$. {
        Provided that the order of the system $n$ is known a priori, an SVD can be utilized to determine the column space of $\mathbf{R}_{22}$:
        \begin{equation}
            \mathbf{R}_{22} = 
            \begin{bmatrix}
                \mathbf{U}_n & \mathbf{U}_n'
            \end{bmatrix}
            \begin{bmatrix}
                \mathbf{S}_n & \mathbf{0}\\
                \mathbf{0} & \mathbf{S_n'}
            \end{bmatrix}
            \mathbf{V}^\intercal,
        \end{equation}where $\mathbf{S}_n$ contains the first $n$ largest singular values of $\mathbf{R}_{22}$ (due to round-off errors {or noise}, there might be more than $n$ nonzero singlular values in $\mathbf{R}_{22}$), $\mathbf{U}_n$ has $n$ columns, and the remaining matrices $\mathbf{U}_n'$, $\mathbf{S}_n$, $\mathbf{S}_n'$, and $\mathbf{V}$ have compatible dimensions. 
        Hence, to obtain an $n$-dimensional observable system realization, we can approximate $\text{span}_{\rm col}(\mathbf{R}_{22}) \approx\text{span}_{\rm col}(\mathbf{U}_{n})$. 
        From $\text{span}_{\rm col}(\mathbf{O}_s(\mathbf A, \mathbf C))=\text{span}_{\rm col}(\mathbf{U}_n)$, a system realization can be obtained as $\mathbf{C}=\mathbf{U}_{n}(1:p,:)$, meaning that $\mathbf{C}$ corresponds to the first $p$ rows of $\mathbf{U}_{n}$, and $\mathbf{A}$ is computed by solving the overdetermined system $\mathbf{U}_{n}(1:(s-1)p,:)\mathbf{A} = \mathbf{U}_{n}(p+1:sp,:)$, which admits a unique solution since $\mathbf{U}_{n}(1:(s-1)p,:)$ has full column rank (note that $s\ge l(\mathbf{A},\mathbf{C})$). Afterwards, ${\mathbf{B}}$ can be obtained by solving a least square problem; see ref.\cite[Theorem 7.1]{verhaegen2007filtering}. Note that once    $(\mathbf{A},\mathbf{C},\mathbf{B})$ corresponds to a system that can generate the input-output data $(\mathbf{U},\mathbf{Y})$, for any $n$-dimensional invertible matrix $\mathbf{T}$, so does $(\mathbf{T}^{-1}\mathbf{A}\mathbf{T},\mathbf{C}\mathbf{T},\mathbf{T}^{-1}\mathbf{B})$. This identification method, also called ``MOESP'' (multivariable output-error state-space), {also applies to the case where $\mathbf{Y}$ is corrupted by i.i.d. measurement noise with zero mean and relatively small variance\cite[page 312]{verhaegen2007filtering}.} 

	}
\noindent \textbf{Constructing the projection $\mathbf{X}^{\rm proj}$ from input-output data.} The projection of $\mathbf{X}$ on the row space of the observability matrix $\mathbf{O}\doteq \mathbf{O}(\mathbf{A},\mathbf{C})$, given by $\mathbf{O}^\dagger\mathbf{O}\mathbf{X}$, plays an important role in the reduced-order functional observer design when only the input-output and functional state data $(\mathbf{U},\mathbf{Y},\mathbf{Z})$ are available.
According to Proposition 2 in Supplementary Note 4, under some standard assumptions, $\text{span}_{\text{{row}}}\left( \mathbf{O}_k^\dagger \mathbf{O}_k {\mathbf{X}}_{k,N} \right) = \text{span}_{\text{{row}}}(\mathbf{V}_p) \cap \text{span}_{\text{{row}}}(\mathbf{V}_f),$ where $k\ge l(\mathbf{A},\mathbf{C})$, $N$ is sufficiently large, $\mathbf{V}_p = \begin{bmatrix} \mathbf{U}_{0,k,N} \\ \mathbf{Y}_{0,k,N} \end{bmatrix}$, $\mathbf{V}_f = \begin{bmatrix} \mathbf{U}_{k,k,N} \\ \mathbf{Y}_{k,k,N} \end{bmatrix}$,  and $\mathbf{O}_k=\mathbf{O}_k(\mathbf A, \mathbf C)$. {We adopt an SVD-based construction for the intersection of two row spaces\cite{moonen1989onandoff}.  To be specific, let the SVD of  $[\mathbf{V}_p^\intercal,\mathbf{V}_f^\intercal]^\intercal$ be
\begin{equation*}
    \begin{bmatrix}
        \mathbf{V}_p \\
        \mathbf{V}_f
    \end{bmatrix} = 
    \begin{bmatrix}
        \mathbf{U}_{11} & \mathbf{U}_{12} \\
        \mathbf{U}_{21} & \mathbf{U}_{22}
    \end{bmatrix}
    \begin{bmatrix}
        \mathbf{S}_{11} & \mathbf{0}\\
        \mathbf{0} & \mathbf{0}
    \end{bmatrix}
    \mathbf{V}^\intercal,
\end{equation*}where $\mathbf{U}_{11},\mathbf{U}_{12},\mathbf{U}_{21},\mathbf{U}_{22},\mathbf{S}_{11}$, and $\mathbf{V}$ are matrices with compatible dimensions. Let the SVD of $\mathbf{U}_{12}^\intercal\cdot \mathbf{U}_{11} \cdot \mathbf{S}_{11}$ be
\begin{equation*}
    \mathbf{U}_{12}^\intercal\cdot \mathbf{U}_{11} \cdot \mathbf{S}_{11} = 
    \begin{bmatrix}
        \mathbf{U}_q & \mathbf{U}_q^\perp
    \end{bmatrix}
    \begin{bmatrix}
        \mathbf{S}_q & \mathbf{0}\\
        \mathbf{0} & \mathbf{0}
    \end{bmatrix}
    \mathbf{V}_q^\intercal,
\end{equation*}where $\mathbf{U}_q$ has the same number of columns as that of $\mathbf{S}_q$.
Then, a basis of 
$\text{span}_{\text{{row}}}( \mathbf{O}_k^\dagger \mathbf{O}_k {\mathbf{X}}_{k,N})$, denoted by ${\mathbf X}^{\rm proj}$, can be constructed as $ {\mathbf X}^{\rm proj}  = 
    \mathbf{U}_q^\intercal  \mathbf{U}_{12}^\intercal \mathbf{V}_p$.
Afterwards, let the data matrix $\mathbf{W}=\mathbf{X}^{\rm proj}$, accordingly, $\mathbf{W}_p=\mathbf{X}^{\rm proj}_{0,N-1}$ and 
$\mathbf{W}_f=\mathbf{X}^{\rm proj}_{1,N-1}$. Then, a reduced-order functional observer can be constructed based on $\mathbf{W}$ in the way described previously. }

\noindent {{\textbf{Generation of complex dynamical networks.}} For the generation of the complex network systems used in Figs.~2-3, we consider random dynamical networks with each node being a two-dimensional subsystem. {The dynamics of each node is governed by the following equation
\begin{equation*}
\begin{array}{l}
\mathbf{x}_i(t+1)=\mathbf{A_i} \mathbf{x}_i(t)+\mathbf{B} v_{i}(t),\\
v_{i}(t) =\delta_{i}^u u_k(t)+ \sum_{j} w_{ij}\mathbf{H} \mathbf{x}_{j}(t), \\
y_i(t) = \delta_{i}^y\mathbf{C}\mathbf{x}_i(t),
\end{array}
\end{equation*}
where $\mathbf{A}_i =\begin{bmatrix}
    * & *\\
    a_{21}^i & 0
\end{bmatrix}$, $\mathbf{B}=\left[0, 1\right]^{\intercal}$, $\mathbf{C}=\left[1,0\right]$, and $\mathbf H=[0,1]$ denote the state, input, output, and coupling matrices, respectively, with $*$ indicating nonzero entries. To introduce nodal heterogeneity, $a_{21}^i$ assumes a nonzero value with probability $0.5$. The network structure is defined by the adjacency matrix $\mathbf{W}=[w_{ij}]_{i,j=1,...,n}$ of randomly generated ER or BA networks (parameters specified in Figs. 2-3 captions), with $w_{ii}=0$ excluding self-loops. All nonzero entries in $\mathbf{A}_i$ and $\mathbf{W}$ are sampled from the uniform distribution on the interval $(0,1)$. Dedicated inputs are uniformly allocated to subsystem second states, with $\delta_{i}^u=1$ indicating external control and $\delta_{i}^u=0$ otherwise. Similarly, $\delta_i^y=1$ denotes sensor placement at the first state of subsystem $i$. The global state matrix is $\mathbf{A}={\rm diag}\{\mathbf{A}_1,\mathbf{A}_2,...,\mathbf{A}_n\} - \mathbf{W} \otimes \mathbf{M}$, where $\otimes$ is the Kronecker product and $\mathbf{M}=\mathbf B\mathbf H$ ($\mathbf{M}_{22}=1$, zero elsewhere). For numerical stability, each realization $\mathbf A$ is normalized by $1.01\rho(\mathbf{A})$. In this setting, following structural functional observability theory\cite{montanari2022functional,zhang2023functional}, we can make a network realization functionally observable with overwhelming probability by placing sensors such that every functional state variable connects to a sensor, either directly or via paths.
}}

\noindent \textbf{Data-based augmentation procedure for reduced-order functional observers.} Here, we provide an iterative procedure (Algorithm 1) to find a matrix $\mathbf{R}_s$ such that~\eqref{condnew1} and \eqref{condnew1-add} hold. Below, for a linear subspace $\mathcal{P}\subseteq\mathbb R^{m}$, $\mathrm{basis}(\mathcal{P})$ denotes a basis matrix whose rows span $\mathcal{P}$ and $\mathcal{P}^\perp=\{\mathbf v\in\mathbb R^{m}:\mathbf u^\intercal \mathbf v=0,\forall \mathbf u\in\mathcal{P}\}$ denotes its orthogonal complement. For a matrix $\mathbf M$, ${\mathcal R}(\mathbf M)$ denotes its row space and ${\rm row}(\mathbf M)$ takes its row number. By $[\mathbf Q_1,\mathbf Q_2]=\mathrm{nullbasis}[\mathbf M_1^\intercal, \mathbf M_2^\intercal]^\intercal$, we mean $[\mathbf Q_1,\mathbf Q_2]$ forms a basis matrix of the left null space of matrix $[\mathbf M_1^\intercal, \mathbf M_2^\intercal]^\intercal$, where the column number of $\mathbf Q_1$ ($\mathbf Q_2$) equals the row number of $\mathbf M_1$ ($\mathbf M_2$). In Algorithm 1, the computation of basis matrices for null spaces and subspace intersections is implemented via SVD as described previously; see also ref.\cite{fernando2010numerical}. A step-by-step explanation of this algorithm can be found in Supplementary Note 4. 

{\small{
\begin{algorithm} 
\caption{Iterative procedure for finding matrix $\mathbf{R}_s$}
\begin{algorithmic}[1] \label{alg1}
\Require Matrices $\mathbf{U}_p, \mathbf{Y}_p, \mathbf{Y}_f, \mathbf{Z}_p, \mathbf{Z}_f, \mathbf{W}_p, \mathbf{W}_f$ of a functionally observable system
\Ensure  Matrix $\mathbf{R}_s$ satisfying conditions \eqref{condnew1} and \eqref{condnew1-add} (if applicable)
\State Initialize $\mathbf{R}_0 \gets \mathbf{0}_{0 \times {\rm row}(\mathbf W_p)}$, $i \gets 1$
\While{${\rm row}(\mathbf R_{i-1})\le {\rm row}(\mathbf W_p)$}
    \State Form $\mathbf{\Pi}_{i-1} = \begin{bmatrix} \mathbf{U}_p \\ \mathbf{Y}_p \\ \mathbf{Y}_f \\ \mathbf{Z}_p \\ \mathbf{R}_{i-1}\mathbf{W}_p \end{bmatrix}$,
          $\mathbf{\Omega}_{i-1} = \begin{bmatrix} \mathbf{\Pi}_{i-1} \\ \mathbf{Z}_f \\ \mathbf{R}_{i-1}\mathbf{W}_f \end{bmatrix}$,
          $a_{i-1} = \mathrm{rank}(\mathbf{\Pi}_{i-1})$,
          $b_{i-1} = \mathrm{rank}(\mathbf{\Omega}_{i-1})$
    \State Compute null space bases: $\left[ \mathbf{\Phi}_{i1}, \mathbf{\Phi}_{i2} \right] \gets \operatorname{nullbasis}\!\begin{bmatrix}\mathbf{\Omega}_{i-1}\\ \mathbf{W}_f\end{bmatrix}$,
          $[\mathbf{\Theta}_{i1}, \mathbf{\Theta}_{i2}] \gets \operatorname{nullbasis}\!\begin{bmatrix}\mathbf{\Omega}_{i-1}\\ \mathbf{W}_p\end{bmatrix}$,
          $\left[ \mathbf{\Xi}_{i1}, \mathbf{\Xi}_{i2} \right] \gets \operatorname{nullbasis}\!\begin{bmatrix}\mathbf{\Pi}_{i-1}\\ \mathbf{W}_p\end{bmatrix}$
    \State Compute basis matrices $\mathbf{T}_i = \operatorname{basis}\!\big( \mathcal{R}(\mathbf{\Theta}_{i2}) \cap \mathcal{R}(\mathbf{\Phi}_{i2}) \big)$,
          $\mathbf{\Gamma}_i = \operatorname{basis}\!\big( \mathcal{R}(\mathbf{\Xi}_{i2})^\perp \cap \mathcal{R}(\mathbf{T}_i) \big)$
    \State Update $\mathbf{\bar R}_i = \begin{bmatrix}\mathbf{R}_{i-1}\\ \mathbf{\Gamma}_i\end{bmatrix}$,
          $\mathbf{\bar \Pi}_i = \begin{bmatrix}\mathbf{\Pi}_{i-1}\\ \mathbf{\Gamma}_i \mathbf{W}_p\end{bmatrix}$,
          $\mathbf{\bar \Omega}_i = \begin{bmatrix}\mathbf{\Omega}_{i-1}\\ \mathbf{\Gamma}_i \mathbf{W}_p\\ \mathbf{\Gamma}_i \mathbf{W}_f\end{bmatrix}$, $\bar a_i = \mathrm{rank}(\mathbf{\bar \Pi}_i)$, $\bar b_i = \mathrm{rank}(\mathbf{\bar \Omega}_i)$
    \If{$\bar a_i = \bar b_i$ \textbf{ and } condition \eqref{condnew1-add} holds for $\mathbf{R}_s=\mathbf{\bar R}_i$}
        \State $\mathbf{R}_s \gets \mathbf{\bar R}_i$; \textbf{break}
    \EndIf
    \State Compute null space bases:
          $\left[ \mathbf{\bar \Xi}_{i1}, \mathbf{\bar \Xi}_{i2} \right] \gets \operatorname{nullbasis}\!\begin{bmatrix}\mathbf{\bar \Pi}_i\\ \mathbf{W}_p\end{bmatrix}$,
          $\left[ \mathbf{\bar \Theta}_{i1}, \mathbf{\bar \Theta}_{i2} \right] \gets \operatorname{nullbasis}\!\begin{bmatrix}\mathbf{\bar \Omega}_i\\ \mathbf{W}_p\end{bmatrix}$,
          $\left[ \mathbf{\bar \Phi}_{i1}, \mathbf{\bar \Phi}_{i2} \right] \gets \operatorname{nullbasis}\!\begin{bmatrix}\mathbf{\bar \Omega}_i\\ \mathbf{W}_f\end{bmatrix}$
    \State Compute $\mathbf{\bar \Gamma}_i^1 = \operatorname{basis}\!\big(\mathcal{R}(\mathbf{\bar \Xi}_{i2})^\perp \cap \mathcal{R}(\mathbf{\bar \Theta}_{i2})\big)$,
          $\mathbf{\bar \Gamma}_i^2 = \operatorname{basis}\!\big(\mathcal{R}(\mathbf{\bar \Xi}_{i2})^\perp \cap \mathcal{R}(\mathbf{\bar \Phi}_{i2})\big)$,
          $\mathbf{\bar \Gamma}_i = \begin{bmatrix}\mathbf{\bar \Gamma}_i^1 \\ \mathbf{\bar \Gamma}_i^2\end{bmatrix}$
    \State Choose $\mathbf{q}_i$:
          if $\mathbf{\bar \Gamma}_i$ nonempty, take an arbitrary row of $\mathbf{\bar \Gamma}_i$;
          else take an arbitrary row of $\operatorname{basis}(\mathcal{R}(\mathbf{\bar \Xi}_{i2})^\perp)$
    \State Update $\mathbf{R}_i = \begin{bmatrix}\mathbf{R}_{i-1}\\ \mathbf{\Gamma}_i\\ \mathbf{q}_i\end{bmatrix}, {\mathbf R}_{s}\gets {\mathbf R}_i, \quad i \gets i+1$
\EndWhile
\State \Return $\mathbf{R}_s$ 
\end{algorithmic}
\end{algorithm}
}}

In each iteration, Algorithm 1 tries to increase $a_i$ while keeping $b_i$ as long as it is possible (otherwise increase both $a_i$ and $b_i$ by one or two; definitions 
of $a_i$ and $b_i$ are given in Algorithm 1). Therefore, it is expected that this procedure can find a $\mathbf R_s$ with a relatively small row number. In the IO-data based design, i.e., $\mathbf W=\mathbf X^{\rm proj}$, this procedure can achieve an observer order upper bounded by $n-p$, matching the minimum order of a reduced-order Luenberger observer for observable systems (Theorem 3 in Supplementary Note 4).

\noindent \textbf{Multi-area power system dynamics and parameters.} 
For every area $i$ of the power system, as in ref.\cite{ma2014distributed}, the linearized LFC model is presented as follows:
    \begin{equation*} \Delta \dot{f}_i = -\frac{{{D}}_i}{M_i}\Delta f_i + \frac{1}{M_i}(\Delta P_{\text{mech},i} - \Delta P_{\text{tie},i} - \Delta P_{\text{L},i}), \end{equation*}
 \begin{equation*} \Delta \dot{P}_{\text{mech},i} = -\frac{1}{T_{\text{t},i}}\Delta P_{\text{mech},i} + \frac{1}{T_{\text{t},i}}\Delta {X}_{\text{g},i}, \end{equation*}
 \begin{equation*} \Delta \dot{{X}}_{\text{g},i} = -\frac{R_i}{T_{\text{g},i}}\Delta f_i - \frac{1}{T_{\text{g},i}}\Delta {X}_{\text{g},i} + \frac{1}{T_{\text{g},i}}\Delta P_{\text{ref},i},  \end{equation*}
 \begin{equation*} \Delta \dot{P}_{\text{tie},i} = \sum_{j = 1}^{N} 2\pi T_{ij}(\Delta f_i - \Delta f_j),
 \end{equation*}
 \begin{equation*} \text{ACE}_i = \beta_i\Delta f_i + \Delta P_{\text{tie},i},
 \end{equation*}
 where \(\Delta f_i\) is the frequency deviation, \({{D}}_i\) is the generator damping coefficient, \(M_i\) is the moment of inertia, \(\Delta P_{\text{mech},i}\) is the generator output power increment, \(\Delta P_{\text{tie},i}\) is the tie-line power deviation, \(\Delta P_{\text{L},i}\) is the external load disturbance, \(T_{\text{t},i}\) is the steam turbine time constant, \(\Delta \mathbf{X}_{\text{g},i}\) is the governor valve position deviation, \(R_i\) is the droop coefficient, \(T_{\text{g},i}\) is the governor time constant, \(\Delta P_{\text{ref},i}\) is the system input, \(T_{ij}\) is the tie-line synchronous power coefficient, \(\Delta f_j\) is the frequency deviation of the interconnected area \(j\), \(\text{ACE}_i\) is the area control error, and \(\beta_i\) is the frequency deviation factor. The parameter values are taken from ref.\cite{shangguan2021adjustable}. The multi-area LFC state space equation is then described as  
$$
\dot{{\mathbf{x}}}(t) = {\mathbf{A}}{\mathbf{x}}(t) + {\mathbf{B}}{\mathbf{u}}(t) + {\mathbf{F}}{\mathbf{d}}(t), \\
$$
$$
\mathbf{y}(k)=\mathbf{C}\mathbf{x}(k),
$$
where $\mathbf{x}(t)=[
\mathbf{x}_1^{\intercal}(t), \cdots,\mathbf{x}_N^{\intercal}(t)
]^{\intercal}$ with $\mathbf{x}_i(t)=[\Delta f_i,\Delta P_{\text{mech},i}, \Delta \mathbf{X}_{\text{g},i},\Delta P_{\text{tie},i}]^{\intercal}$, $\mathbf{u}(t)$, $\mathbf{y}(t)$ and $\mathbf{d}(t)$ represent the lumped combinations of inputs $\Delta P_{\text{ref},i}$, outputs $\text{ACE}_i$, and load disturbance $\Delta P_{\text{L},i}$ of the $10$ generators, respectively. In the simulation, we use the forward difference method to discretize the model with a time step of $0.01$. The resulting system matrix is normalized by its spectral radius to ensure numerical stability.
In order to achieve stable frequency deviation and interconnection line exchange power under any step load disturbance, the integral of the output ACE is used as the controlled feedback variable \cite{ma2014distributed}. Ultimately, the augmented discrete-time system equations are obtained.

\noindent \textbf{Computational methods and implementation details.} All simulations were performed using MATLAB R2024a on a desktop computer with a 3.40 GHz Intel Core i7 processor and 16 GB of RAM. Pseudoinverses are computed using MATLAB's \texttt{pinv} function with default tolerance settings. Null spaces are determined via SVD with a default threshold of $10^{-5}$ unless otherwise stated. Observer poles are assigned using MATLAB's \texttt{place} command, with all observers designed under identical rightmost pole constraints to ensure consistent decay rates in RRMSE comparisons. Specifically, the rightmost poles of observers are constrained below 0.4 in Fig.~\ref{fig: motivation-framework}, and below 0.9 in Figs.~\ref{fig:fig2} and~\ref{fig:fig3}, where the relaxed constraint accommodates potential infeasibility of dense pole placement for large-dimensional matrices. In all simulations, unless otherwise stated, the initial states of the
original systems are randomly selected, while all observers started from zero. The rank condition $\operatorname{rank}([\mathbf{M_1^\intercal},\mathbf{M_2}^\intercal]^\intercal)=\operatorname{rank}(\mathbf{M_1})$ is verified by checking whether $\|\mathbf M_2(\mathbf I-\mathbf M_1^\dagger \mathbf M_1)\|_F/\|\mathbf M_2\|_F < \epsilon^2$, where $\epsilon=5\times 10^{-7}$ and $\|\cdot\|_F$ denotes the Frobenius norm.

\textbf{Data availability} Data generated or analyzed during this study are included within this paper and its supplementary information files.   
Data of the IEEE-39 New England power grid in Fig. 5 can be found in the corresponding ref.\cite{shangguan2021adjustable}. All data are also available in the public GitHub repository: \url{https://github.com/Yuanzhang2014/DataFunObsv}.

\textbf{Code availability}
The code of this study is freely available in the public GitHub repository: \\~ \url{https://github.com/Yuanzhang2014/DataFunObsv}.

\textbf{Acknowledgements}
	We thank Zhongqi Sun, Yi Yu, and Yu Wang for helpful comments and suggestions. Y.Z, Z.L, W.X, J.W, and R.C gratefully acknowledge the support from the National Natural Science Foundation of China under Grant 62373059.
	


\textbf{Competing interests}
The authors declare no competing interests. 
   {\footnotesize{
	\bibliography{yuanz3.bib}

\begin{thebibliography}{10}
\expandafter\ifx\csname url\endcsname\relax
  \def\url#1{\texttt{#1}}\fi
\expandafter\ifx\csname urlprefix\endcsname\relax\def\urlprefix{URL }\fi
\providecommand{\bibinfo}[2]{#2}
\providecommand{\eprint}[2][]{\url{#2}}

\bibitem{pickard2025dynamic}
\bibinfo{author}{Pickard, J.} \emph{et~al.}
\newblock \bibinfo{title}{Dynamic sensor selection for biomarker discovery}.
\newblock \emph{\bibinfo{journal}{Proceedings of the National Academy of
  Sciences}} \textbf{\bibinfo{volume}{122}}, \bibinfo{pages}{e2501324122}
  (\bibinfo{year}{2025}).

\bibitem{liu2013observability}
\bibinfo{author}{Liu, Y.-Y.}, \bibinfo{author}{Slotine, J.-J.} \&
  \bibinfo{author}{Barab{\'a}si, A.-L.}
\newblock \bibinfo{title}{Observability of complex systems}.
\newblock \emph{\bibinfo{journal}{Proceedings of the National Academy of
  Sciences}} \textbf{\bibinfo{volume}{110}}, \bibinfo{pages}{2460--2465}
  (\bibinfo{year}{2013}).

\bibitem{maclaren2025observing}
\bibinfo{author}{MacLaren, N.~G.}, \bibinfo{author}{Barzel, B.} \&
  \bibinfo{author}{Masuda, N.}
\newblock \bibinfo{title}{Observing network dynamics through sentinel nodes}.
\newblock \emph{\bibinfo{journal}{Nature Communications}}
  \textbf{\bibinfo{volume}{16}}, \bibinfo{pages}{10211} (\bibinfo{year}{2025}).

\bibitem{simon2006optimal}
\bibinfo{author}{Simon, D.}
\newblock \emph{\bibinfo{title}{Optimal State Estimation: Kalman, H infinity,
  and Nonlinear Approaches}} (\bibinfo{publisher}{John Wiley \& Sons},
  \bibinfo{year}{2006}).

\bibitem{chen2012robust}
\bibinfo{author}{Chen, J.} \& \bibinfo{author}{Patton, R.~J.}
\newblock \emph{\bibinfo{title}{Robust Model-Based Fault Diagnosis for Dynamic
  Systems}}, vol.~\bibinfo{volume}{3} (\bibinfo{publisher}{Springer Science \&
  Business Media}, \bibinfo{year}{2012}).

\bibitem{Fixed_Mode}
\bibinfo{author}{Wang, S.~H.} \& \bibinfo{author}{Davison, E.~J.}
\newblock \bibinfo{title}{On the stabilization of decentralized control
  systems}.
\newblock \emph{\bibinfo{journal}{IEEE Transactions on Automatic Control}}
  \textbf{\bibinfo{volume}{18}}, \bibinfo{pages}{473--478}
  (\bibinfo{year}{1973}).

\bibitem{Stefano2012Stability}
\bibinfo{author}{Allesina, S.} \& \bibinfo{author}{Tang, S.}
\newblock \bibinfo{title}{Stability criteria for complex ecosystems}.
\newblock \emph{\bibinfo{journal}{Nature}} \textbf{\bibinfo{volume}{483}},
  \bibinfo{pages}{205--208} (\bibinfo{year}{2012}).

\bibitem{luenberger1966observers}
\bibinfo{author}{Luenberger, D.}
\newblock \bibinfo{title}{Observers for multivariable systems}.
\newblock \emph{\bibinfo{journal}{IEEE Transactions on Automatic Control}}
  \textbf{\bibinfo{volume}{11}}, \bibinfo{pages}{190--197}
  (\bibinfo{year}{1966}).

\bibitem{luenberger1971introduction}
\bibinfo{author}{Luenberger, D.}
\newblock \bibinfo{title}{An introduction to observers}.
\newblock \emph{\bibinfo{journal}{IEEE Transactions on Automatic Control}}
  \textbf{\bibinfo{volume}{16}}, \bibinfo{pages}{596--602}
  (\bibinfo{year}{1971}).

\bibitem{montanari2022functional}
\bibinfo{author}{Montanari, A.~N.}, \bibinfo{author}{Duan, C.},
  \bibinfo{author}{Aguirre, L.~A.} \& \bibinfo{author}{Motter, A.~E.}
\newblock \bibinfo{title}{Functional observability and target state estimation
  in large-scale networks}.
\newblock \emph{\bibinfo{journal}{Proceedings of the National Academy of
  Sciences (PNAS)}} \textbf{\bibinfo{volume}{119}},
  \bibinfo{pages}{e2113750119} (\bibinfo{year}{2022}).

\bibitem{darouach2000existence}
\bibinfo{author}{Darouach, M.}
\newblock \bibinfo{title}{Existence and design of functional observers for
  linear systems}.
\newblock \emph{\bibinfo{journal}{IEEE Transactions on Automatic Control}}
  \textbf{\bibinfo{volume}{45}}, \bibinfo{pages}{940--943}
  (\bibinfo{year}{2000}).

\bibitem{fernando2010functional}
\bibinfo{author}{Fernando, T.~L.}, \bibinfo{author}{Trinh, H.~M.} \&
  \bibinfo{author}{Jennings, L.}
\newblock \bibinfo{title}{Functional observability and the design of minimum
  order linear functional observers}.
\newblock \emph{\bibinfo{journal}{IEEE Transactions on Automatic Control}}
  \textbf{\bibinfo{volume}{55}}, \bibinfo{pages}{1268--1273}
  (\bibinfo{year}{2010}).

\bibitem{gao2014target}
\bibinfo{author}{Gao, J.}, \bibinfo{author}{Liu, Y.-Y.},
  \bibinfo{author}{D'souza, R.~M.} \& \bibinfo{author}{Barab{\'a}si, A.-L.}
\newblock \bibinfo{title}{Target control of complex networks}.
\newblock \emph{\bibinfo{journal}{Nature Communications}}
  \textbf{\bibinfo{volume}{5}}, \bibinfo{pages}{1--8} (\bibinfo{year}{2014}).

\bibitem{klickstein2017energy}
\bibinfo{author}{Klickstein, I.}, \bibinfo{author}{Shirin, A.} \&
  \bibinfo{author}{Sorrentino, F.}
\newblock \bibinfo{title}{Energy scaling of targeted optimal control of complex
  networks}.
\newblock \emph{\bibinfo{journal}{Nature Communications}}
  \textbf{\bibinfo{volume}{8}}, \bibinfo{pages}{15145} (\bibinfo{year}{2017}).

\bibitem{yabe2022toward}
\bibinfo{author}{Yabe, T.}, \bibinfo{author}{Rao, P. S.~C.},
  \bibinfo{author}{Ukkusuri, S.~V.} \& \bibinfo{author}{Cutter, S.~L.}
\newblock \bibinfo{title}{Toward data-driven, dynamical complex systems
  approaches to disaster resilience}.
\newblock \emph{\bibinfo{journal}{Proceedings of the National Academy of
  Sciences}} \textbf{\bibinfo{volume}{119}}, \bibinfo{pages}{e2111997119}
  (\bibinfo{year}{2022}).

\bibitem{turk2013functional}
\bibinfo{author}{Turk-Browne, N.~B.}
\newblock \bibinfo{title}{Functional interactions as big data in the human
  brain}.
\newblock \emph{\bibinfo{journal}{Science}} \textbf{\bibinfo{volume}{342}},
  \bibinfo{pages}{580--584} (\bibinfo{year}{2013}).

\bibitem{ushio2018fluctuating}
\bibinfo{author}{Ushio, M.} \emph{et~al.}
\newblock \bibinfo{title}{Fluctuating interaction network and time-varying
  stability of a natural fish community}.
\newblock \emph{\bibinfo{journal}{Nature}} \textbf{\bibinfo{volume}{554}},
  \bibinfo{pages}{360--363} (\bibinfo{year}{2018}).

\bibitem{raissi2020hidden}
\bibinfo{author}{Raissi, M.}, \bibinfo{author}{Yazdani, A.} \&
  \bibinfo{author}{Karniadakis, G.~E.}
\newblock \bibinfo{title}{Hidden fluid mechanics: Learning velocity and
  pressure fields from flow visualizations}.
\newblock \emph{\bibinfo{journal}{Science}} \textbf{\bibinfo{volume}{367}},
  \bibinfo{pages}{1026--1030} (\bibinfo{year}{2020}).

\bibitem{marx2013big}
\bibinfo{author}{Marx, V.}
\newblock \bibinfo{title}{The big challenges of big data}.
\newblock \emph{\bibinfo{journal}{Nature}} \textbf{\bibinfo{volume}{498}},
  \bibinfo{pages}{255--260} (\bibinfo{year}{2013}).

\bibitem{de2019formulas}
\bibinfo{author}{De~Persis, C.} \& \bibinfo{author}{Tesi, P.}
\newblock \bibinfo{title}{Formulas for data-driven control: Stabilization,
  optimality, and robustness}.
\newblock \emph{\bibinfo{journal}{IEEE Transactions on Automatic Control}}
  \textbf{\bibinfo{volume}{65}}, \bibinfo{pages}{909--924}
  (\bibinfo{year}{2020}).

\bibitem{baggio2021data}
\bibinfo{author}{Baggio, G.}, \bibinfo{author}{Bassett, D.~S.} \&
  \bibinfo{author}{Pasqualetti, F.}
\newblock \bibinfo{title}{Data-driven control of complex networks}.
\newblock \emph{\bibinfo{journal}{Nature Communications}}
  \textbf{\bibinfo{volume}{12}}, \bibinfo{pages}{1429} (\bibinfo{year}{2021}).

\bibitem{yeung2019learning}
\bibinfo{author}{Yeung, E.}, \bibinfo{author}{Kundu, S.} \&
  \bibinfo{author}{Hodas, N.}
\newblock \bibinfo{title}{Learning deep neural network representations for
  koopman operators of nonlinear dynamical systems}.
\newblock In \emph{\bibinfo{booktitle}{2019 American Control Conference
  (ACC)}}, \bibinfo{pages}{4832--4839} (\bibinfo{organization}{IEEE},
  \bibinfo{year}{2019}).

\bibitem{coulson2019data}
\bibinfo{author}{Coulson, J.}, \bibinfo{author}{Lygeros, J.} \&
  \bibinfo{author}{D{\"o}rfler, F.}
\newblock \bibinfo{title}{Data-enabled predictive control: In the shallows of
  the deepc}.
\newblock In \emph{\bibinfo{booktitle}{2019 18th European control conference
  (ECC)}}, \bibinfo{pages}{307--312} (\bibinfo{organization}{IEEE},
  \bibinfo{year}{2019}).

\bibitem{gevers1993towards}
\bibinfo{author}{Gevers, M.}
\newblock \bibinfo{title}{Towards a joint design of identification and
  control?}
\newblock In \emph{\bibinfo{booktitle}{Essays on Control: Perspectives in the
  Theory and its Applications}}, \bibinfo{pages}{111--151}
  (\bibinfo{publisher}{Springer}, \bibinfo{year}{1993}).

\bibitem{van1995identification}
\bibinfo{author}{Van Den~Hof, P.~M.} \& \bibinfo{author}{Schrama, R.~J.}
\newblock \bibinfo{title}{Identification and control—closed-loop issues}.
\newblock \emph{\bibinfo{journal}{Automatica}} \textbf{\bibinfo{volume}{31}},
  \bibinfo{pages}{1751--1770} (\bibinfo{year}{1995}).

\bibitem{champion2019data}
\bibinfo{author}{Champion, K.}, \bibinfo{author}{Lusch, B.},
  \bibinfo{author}{Kutz, J.~N.} \& \bibinfo{author}{Brunton, S.~L.}
\newblock \bibinfo{title}{Data-driven discovery of coordinates and governing
  equations}.
\newblock \emph{\bibinfo{journal}{Proceedings of the National Academy of
  Sciences}} \textbf{\bibinfo{volume}{116}}, \bibinfo{pages}{22445--22451}
  (\bibinfo{year}{2019}).

\bibitem{van2020data}
\bibinfo{author}{Van~Waarde, H.~J.}, \bibinfo{author}{Eising, J.},
  \bibinfo{author}{Trentelman, H.~L.} \& \bibinfo{author}{Camlibel, M.~K.}
\newblock \bibinfo{title}{Data informativity: A new perspective on data-driven
  analysis and control}.
\newblock \emph{\bibinfo{journal}{IEEE Transactions on Automatic Control}}
  \textbf{\bibinfo{volume}{65}}, \bibinfo{pages}{4753--4768}
  (\bibinfo{year}{2020}).

\bibitem{berberich2020data}
\bibinfo{author}{Berberich, J.}, \bibinfo{author}{K{\"o}hler, J.},
  \bibinfo{author}{M{\"u}ller, M.~A.} \& \bibinfo{author}{Allg{\"o}wer, F.}
\newblock \bibinfo{title}{Data-driven model predictive control with stability
  and robustness guarantees}.
\newblock \emph{\bibinfo{journal}{IEEE Transactions on Automatic Control}}
  \textbf{\bibinfo{volume}{66}}, \bibinfo{pages}{1702--1717}
  (\bibinfo{year}{2020}).

\bibitem{krishnan2021direct}
\bibinfo{author}{Krishnan, V.} \& \bibinfo{author}{Pasqualetti, F.}
\newblock \bibinfo{title}{On direct vs indirect data-driven predictive
  control}.
\newblock In \emph{\bibinfo{booktitle}{2021 60th IEEE Conference on Decision
  and Control (CDC)}}, \bibinfo{pages}{736--741} (\bibinfo{organization}{IEEE},
  \bibinfo{year}{2021}).

\bibitem{breschi2023data}
\bibinfo{author}{Breschi, V.}, \bibinfo{author}{Chiuso, A.} \&
  \bibinfo{author}{Formentin, S.}
\newblock \bibinfo{title}{Data-driven predictive control in a stochastic
  setting: a unified framework}.
\newblock \emph{\bibinfo{journal}{Automatica}} \textbf{\bibinfo{volume}{152}},
  \bibinfo{pages}{110961} (\bibinfo{year}{2023}).

\bibitem{markovsky2008data}
\bibinfo{author}{Markovsky, I.} \& \bibinfo{author}{Rapisarda, P.}
\newblock \bibinfo{title}{Data-driven simulation and control}.
\newblock \emph{\bibinfo{journal}{International Journal of Control}}
  \textbf{\bibinfo{volume}{81}}, \bibinfo{pages}{1946--1959}
  (\bibinfo{year}{2008}).

\bibitem{wu2024predicting}
\bibinfo{author}{Wu, T.} \emph{et~al.}
\newblock \bibinfo{title}{Predicting multiple observations in complex systems
  through low-dimensional embeddings}.
\newblock \emph{\bibinfo{journal}{Nature Communications}}
  \textbf{\bibinfo{volume}{15}}, \bibinfo{pages}{2242} (\bibinfo{year}{2024}).

\bibitem{lusch2018deep}
\bibinfo{author}{Lusch, B.}, \bibinfo{author}{Kutz, J.~N.} \&
  \bibinfo{author}{Brunton, S.~L.}
\newblock \bibinfo{title}{Deep learning for universal linear embeddings of
  nonlinear dynamics}.
\newblock \emph{\bibinfo{journal}{Nature Communications}}
  \textbf{\bibinfo{volume}{9}}, \bibinfo{pages}{4950} (\bibinfo{year}{2018}).

\bibitem{chang2019neural}
\bibinfo{author}{Chang, Y.-C.}, \bibinfo{author}{Roohi, N.} \&
  \bibinfo{author}{Gao, S.}
\newblock \bibinfo{title}{Neural lyapunov control}.
\newblock \emph{\bibinfo{journal}{Advances in neural information processing
  systems}} \textbf{\bibinfo{volume}{32}} (\bibinfo{year}{2019}).

\bibitem{Turan2021}
\bibinfo{author}{Turan, M.~S.} \& \bibinfo{author}{Ferrari-Trecate, G.}
\newblock \bibinfo{title}{Data-driven unknown-input observers and state
  estimation}.
\newblock \emph{\bibinfo{journal}{IEEE Control Systems Letters}}
  \textbf{\bibinfo{volume}{6}}, \bibinfo{pages}{1424--1429}
  (\bibinfo{year}{2021}).

\bibitem{wolff2024robust}
\bibinfo{author}{Wolff, T.~M.}, \bibinfo{author}{Lopez, V.~G.} \&
  \bibinfo{author}{M{\"u}ller, M.~A.}
\newblock \bibinfo{title}{Robust data-driven moving horizon estimation for
  linear discrete-time systems}.
\newblock \emph{\bibinfo{journal}{IEEE Transactions on Automatic Control}}
  \textbf{\bibinfo{volume}{69}}, \bibinfo{pages}{5598--5604}
  (\bibinfo{year}{2024}).

\bibitem{disaro2024equivalence}
\bibinfo{author}{Disar{\`o}, G.} \& \bibinfo{author}{Valcher, M.~E.}
\newblock \bibinfo{title}{On the equivalence of model-based and data-driven
  approaches to the design of unknown-input observers}.
\newblock \emph{\bibinfo{journal}{IEEE Transactions on Automatic Control}}
  (\bibinfo{year}{2024}).

\bibitem{niazi2020average}
\bibinfo{author}{Niazi, M. U.~B.}, \bibinfo{author}{Canudas-de Wit, C.} \&
  \bibinfo{author}{Kibangou, A.~Y.}
\newblock \bibinfo{title}{Average state estimation in large-scale clustered
  network systems}.
\newblock \emph{\bibinfo{journal}{IEEE Transactions on Control of Network
  Systems}} \textbf{\bibinfo{volume}{7}}, \bibinfo{pages}{1736--1745}
  (\bibinfo{year}{2020}).

\bibitem{trinh2011functional}
\bibinfo{author}{Trinh, H.} \& \bibinfo{author}{Fernando, T.}
\newblock \emph{\bibinfo{title}{Functional Observers for Dynamical Systems}},
  vol. \bibinfo{volume}{420} (\bibinfo{publisher}{Springer Science \& Business
  Media}, \bibinfo{year}{2011}).

\bibitem{fernando2010functional2}
\bibinfo{author}{Fernando, T.}, \bibinfo{author}{Jennings, L.} \&
  \bibinfo{author}{Trinh, H.}
\newblock \bibinfo{title}{Functional observability}.
\newblock In \emph{\bibinfo{booktitle}{2010 Fifth International Conference on
  Information and Automation for Sustainability}}, \bibinfo{pages}{421--423}
  (\bibinfo{organization}{IEEE}, \bibinfo{year}{2010}).

\bibitem{markovsky2023data}
\bibinfo{author}{Markovsky, I.}, \bibinfo{author}{Huang, L.} \&
  \bibinfo{author}{D{\"o}rfler, F.}
\newblock \bibinfo{title}{Data-driven control based on the behavioral approach:
  From theory to applications in power systems}.
\newblock \emph{\bibinfo{journal}{IEEE Control Systems Magazine}}
  \textbf{\bibinfo{volume}{43}}, \bibinfo{pages}{28--68}
  (\bibinfo{year}{2023}).

\bibitem{verhaegen2007filtering}
\bibinfo{author}{Verhaegen, M.} \& \bibinfo{author}{Verdult, V.}
\newblock \emph{\bibinfo{title}{Filtering and System Identification: a Least
  Squares Approach}} (\bibinfo{publisher}{Cambridge University Press},
  \bibinfo{year}{2007}).

\bibitem{Y.Y.2011Controllability}
\bibinfo{author}{Liu, Y.~Y.}, \bibinfo{author}{Slotine, J.~J.} \&
  \bibinfo{author}{Barabasi, A.~L.}
\newblock \bibinfo{title}{Controllability of complex networks}.
\newblock \emph{\bibinfo{journal}{Nature}} \textbf{\bibinfo{volume}{48}},
  \bibinfo{pages}{167--173} (\bibinfo{year}{2011}).

\bibitem{Yuan2013Exact}
\bibinfo{author}{Yuan, Z.}, \bibinfo{author}{Zhao, C.}, \bibinfo{author}{Di,
  Z.}, \bibinfo{author}{Wang, W.~X.} \& \bibinfo{author}{Lai, Y.~C.}
\newblock \bibinfo{title}{Exact controllability of complex networks.}
\newblock \emph{\bibinfo{journal}{Nature Communications}}
  \textbf{\bibinfo{volume}{4}}, \bibinfo{pages}{2447} (\bibinfo{year}{2013}).

\bibitem{zhang2023functional}
\bibinfo{author}{Zhang, Y.}, \bibinfo{author}{Fernando, T.} \&
  \bibinfo{author}{Darouach, M.}
\newblock \bibinfo{title}{Functional observability, structural functional
  observability, and optimal sensor placement}.
\newblock \emph{\bibinfo{journal}{IEEE Transactions on Automatic Control}}
  \textbf{\bibinfo{volume}{70}}, \bibinfo{pages}{1592--1607}
  (\bibinfo{year}{2025}).

\bibitem{koopman1931hamiltonian}
\bibinfo{author}{Koopman, B.~O.}
\newblock \bibinfo{title}{Hamiltonian systems and transformation in hilbert
  space}.
\newblock \emph{\bibinfo{journal}{Proceedings of the National Academy of
  Sciences}} \textbf{\bibinfo{volume}{17}}, \bibinfo{pages}{315--318}
  (\bibinfo{year}{1931}).

\bibitem{shang2024willems}
\bibinfo{author}{Shang, X.}, \bibinfo{author}{Cort{\'e}s, J.} \&
  \bibinfo{author}{Zheng, Y.}
\newblock \bibinfo{title}{Willems’ fundamental lemma for nonlinear systems
  with koopman linear embedding}.
\newblock \emph{\bibinfo{journal}{IEEE Control Systems Letters}}
  (\bibinfo{year}{2024}).

\bibitem{williams2015data}
\bibinfo{author}{Williams, M.~O.}, \bibinfo{author}{Kevrekidis, I.~G.} \&
  \bibinfo{author}{Rowley, C.~W.}
\newblock \bibinfo{title}{A data--driven approximation of the koopman operator:
  Extending dynamic mode decomposition}.
\newblock \emph{\bibinfo{journal}{Journal of Nonlinear Science}}
  \textbf{\bibinfo{volume}{25}}, \bibinfo{pages}{1307--1346}
  (\bibinfo{year}{2015}).

\bibitem{surana2016linear}
\bibinfo{author}{Surana, A.} \& \bibinfo{author}{Banaszuk, A.}
\newblock \bibinfo{title}{Linear observer synthesis for nonlinear systems using
  koopman operator framework}.
\newblock \emph{\bibinfo{journal}{IFAC-PapersOnLine}}
  \textbf{\bibinfo{volume}{49}}, \bibinfo{pages}{716--723}
  (\bibinfo{year}{2016}).

\bibitem{rossman1994epanet}
\bibinfo{author}{Rossman, L.~A.} \emph{et~al.}
\newblock \bibinfo{title}{Epanet users manual}  (\bibinfo{year}{1994}).

\bibitem{burgschweiger2009optimization}
\bibinfo{author}{Burgschweiger, J.}, \bibinfo{author}{Gn{\"a}dig, B.} \&
  \bibinfo{author}{Steinbach, M.~C.}
\newblock \bibinfo{title}{Optimization models for operative planning in
  drinking water networks}.
\newblock \emph{\bibinfo{journal}{Optimization and Engineering}}
  \textbf{\bibinfo{volume}{10}}, \bibinfo{pages}{43--73}
  (\bibinfo{year}{2009}).

\bibitem{boulos2006comprehensive}
\bibinfo{author}{Boulos, P.~F.}, \bibinfo{author}{Lansey, K.~E.} \&
  \bibinfo{author}{Karney, B.~W.}
\newblock \emph{\bibinfo{title}{Comprehensive water distribution systems
  analysis handbook for engineers and planners}} (\bibinfo{publisher}{American
  Water Works Association}, \bibinfo{year}{2006}).

\bibitem{brown2002history}
\bibinfo{author}{Brown, G.~O.}
\newblock \bibinfo{title}{The history of the darcy-weisbach equation for pipe
  flow resistance}.
\newblock In \emph{\bibinfo{booktitle}{Environmental and water resources
  history}}, \bibinfo{pages}{34--43} (\bibinfo{year}{2002}).

\bibitem{fawzi2014secure}
\bibinfo{author}{Fawzi, H.}, \bibinfo{author}{Tabuada, P.} \&
  \bibinfo{author}{Diggavi, S.}
\newblock \bibinfo{title}{Secure estimation and control for cyber-physical
  systems under adversarial attacks}.
\newblock \emph{\bibinfo{journal}{IEEE Transactions on Automatic Control}}
  \textbf{\bibinfo{volume}{59}}, \bibinfo{pages}{1454--1467}
  (\bibinfo{year}{2014}).

\bibitem{pan2002adaptive}
\bibinfo{author}{Pan, C.-T.} \& \bibinfo{author}{Liaw, C.-M.}
\newblock \bibinfo{title}{An adaptive controller for power system
  load-frequency control}.
\newblock \emph{\bibinfo{journal}{IEEE Transactions on Power Systems}}
  \textbf{\bibinfo{volume}{4}}, \bibinfo{pages}{122--128}
  (\bibinfo{year}{2002}).

\bibitem{shangguan2021adjustable}
\bibinfo{author}{Shangguan, X.-C.}, \bibinfo{author}{He, Y.},
  \bibinfo{author}{Zhang, C.-K.}, \bibinfo{author}{Jiang, L.} \&
  \bibinfo{author}{Wu, M.}
\newblock \bibinfo{title}{Adjustable event-triggered load frequency control of
  power systems using control-performance-standard-based fuzzy logic}.
\newblock \emph{\bibinfo{journal}{IEEE Transactions on Fuzzy Systems}}
  \textbf{\bibinfo{volume}{30}}, \bibinfo{pages}{3297--3311}
  (\bibinfo{year}{2021}).

\bibitem{ma2014distributed}
\bibinfo{author}{Ma, M.}, \bibinfo{author}{Chen, H.}, \bibinfo{author}{Liu, X.}
  \& \bibinfo{author}{Allg{\"o}wer, F.}
\newblock \bibinfo{title}{Distributed model predictive load frequency control
  of multi-area interconnected power system}.
\newblock \emph{\bibinfo{journal}{International Journal of Electrical Power \&
  Energy Systems}} \textbf{\bibinfo{volume}{62}}, \bibinfo{pages}{289--298}
  (\bibinfo{year}{2014}).

\bibitem{montanari2025duality}
\bibinfo{author}{Montanari, A.~N.}, \bibinfo{author}{Duan, C.} \&
  \bibinfo{author}{Motter, A.~E.}
\newblock \bibinfo{title}{Duality between controllability and observability for
  target control and estimation in networks}.
\newblock \emph{\bibinfo{journal}{IEEE Transactions on Automatic Control}}
  \textbf{\bibinfo{volume}{70}}, \bibinfo{pages}{5584--5591}
  (\bibinfo{year}{2025}).

\bibitem{hindmarsh1984model}
\bibinfo{author}{Hindmarsh, J.~L.} \& \bibinfo{author}{Rose, R.}
\newblock \bibinfo{title}{A model of neuronal bursting using three coupled
  first order differential equations}.
\newblock \emph{\bibinfo{journal}{Proceedings of the Royal society of London.
  Series B. Biological sciences}} \textbf{\bibinfo{volume}{221}},
  \bibinfo{pages}{87--102} (\bibinfo{year}{1984}).

\bibitem{korda2018linear}
\bibinfo{author}{Korda, M.} \& \bibinfo{author}{Mezi{\'c}, I.}
\newblock \bibinfo{title}{Linear predictors for nonlinear dynamical systems:
  Koopman operator meets model predictive control}.
\newblock \emph{\bibinfo{journal}{Automatica}} \textbf{\bibinfo{volume}{93}},
  \bibinfo{pages}{149--160} (\bibinfo{year}{2018}).

\bibitem{Brunton2016SINDy}
\bibinfo{author}{Brunton, S.~L.}, \bibinfo{author}{Proctor, J.~L.} \&
  \bibinfo{author}{Kutz, J.~N.}
\newblock \bibinfo{title}{Discovering governing equations from data by sparse
  identification of nonlinear dynamical systems}.
\newblock \emph{\bibinfo{journal}{Proceedings of the National Academy of
  Sciences}} \textbf{\bibinfo{volume}{113}}, \bibinfo{pages}{3932--3937}
  (\bibinfo{year}{2016}).

\bibitem{montanari2022functional-nonlinear}
\bibinfo{author}{Montanari, A.~N.}, \bibinfo{author}{Freitas, L.},
  \bibinfo{author}{Proverbio, D.} \& \bibinfo{author}{Gon{\c{c}}alves, J.}
\newblock \bibinfo{title}{Functional observability and subspace reconstruction
  in nonlinear systems}.
\newblock \emph{\bibinfo{journal}{Physical Review Research}}
  \textbf{\bibinfo{volume}{4}}, \bibinfo{pages}{043195} (\bibinfo{year}{2022}).

\bibitem{bertipaglia2024unscented}
\bibinfo{author}{Bertipaglia, A.}, \bibinfo{author}{Alirezaei, M.},
  \bibinfo{author}{Happee, R.} \& \bibinfo{author}{Shyrokau, B.}
\newblock \bibinfo{title}{An unscented kalman filter-informed neural network
  for vehicle sideslip angle estimation}.
\newblock \emph{\bibinfo{journal}{IEEE Transactions on Vehicular Technology}}
  \textbf{\bibinfo{volume}{73}}, \bibinfo{pages}{12731--12746}
  (\bibinfo{year}{2024}).

\bibitem{morato2024data}
\bibinfo{author}{Morato, M.~M.} \& \bibinfo{author}{Felix, M.~S.}
\newblock \bibinfo{title}{Data science and model predictive control:: A survey
  of recent advances on data-driven mpc algorithms}.
\newblock \emph{\bibinfo{journal}{Journal of Process Control}}
  \textbf{\bibinfo{volume}{144}}, \bibinfo{pages}{103327}
  (\bibinfo{year}{2024}).

\bibitem{fernando2025existence}
\bibinfo{author}{Fernando, T.} \& \bibinfo{author}{Darouach, M.}
\newblock \bibinfo{title}{Existence and design of target output controllers}.
\newblock \emph{\bibinfo{journal}{IEEE Transactions on Automatic Control}}
  \textbf{\bibinfo{volume}{70}}, \bibinfo{pages}{6104--6110}
  (\bibinfo{year}{2025}).

\bibitem{Darouach2023FunctionalDA}
\bibinfo{author}{Darouach, M.} \& \bibinfo{author}{Fernando, T.~L.}
\newblock \bibinfo{title}{Functional detectability and asymptotic functional
  observer design}.
\newblock \emph{\bibinfo{journal}{IEEE Transactions on Automatic Control}}
  \textbf{\bibinfo{volume}{68}}, \bibinfo{pages}{975--990}
  (\bibinfo{year}{2023}).

\bibitem{verahegen1992subspace}
\bibinfo{author}{Verahegen, M.} \& \bibinfo{author}{Dewilde, P.}
\newblock \bibinfo{title}{Subspace model identification. part i: The
  output-error state-space model identification class of algorithm}.
\newblock \emph{\bibinfo{journal}{International Journal of Control}}
  \textbf{\bibinfo{volume}{56}}, \bibinfo{pages}{1187--1210}
  (\bibinfo{year}{1992}).

\bibitem{moonen1989onandoff}
\bibinfo{author}{Moonen, M.}, \bibinfo{author}{De~Moor, B.},
  \bibinfo{author}{Vandenberghe, L.} \& \bibinfo{author}{Vandewalle, J.}
\newblock \bibinfo{title}{On-and off-line identification of linear state-space
  models}.
\newblock \emph{\bibinfo{journal}{International Journal of Control}}
  \textbf{\bibinfo{volume}{49}}, \bibinfo{pages}{219--232}
  (\bibinfo{year}{1989}).

\bibitem{fernando2010numerical}
\bibinfo{author}{Fernando, T.}, \bibinfo{author}{Jennings, L.} \&
  \bibinfo{author}{Trinh, H.}
\newblock \bibinfo{title}{Numerical implementation of a functional
  observability algorithm: A singular value decomposition approach}.
\newblock In \emph{\bibinfo{booktitle}{2010 IEEE Asia Pacific Conference on
  Circuits and Systems}}, \bibinfo{pages}{796--799}
  (\bibinfo{organization}{IEEE}, \bibinfo{year}{2010}).

\end{thebibliography}
		}}
        
\end{document}